\documentclass[12pt]{article}
\usepackage[margin=1in]{geometry}
\usepackage[round]{natbib}
\usepackage{amsmath,amssymb,authblk,graphicx}

\DeclareMathOperator*{\argmax}{arg\,max}
\DeclareMathOperator*{\argmin}{arg\,min}

\def\bdelta{\boldsymbol{\delta}}
\def\E{\mathrm{E}}

\def\bt{\mathbf{t}}
\def\bu{\mathbf{u}}
\def\bmu{\boldsymbol{\mu}}

\def\bw{\mathbf{w}}
\def\bomega{\boldsymbol{\omega}}
\def\bX{\mathbf{X}}
\def\btheta{\boldsymbol{\theta}}

\newtheorem{lemma}{Lemma}
\newtheorem{theorem}{Theorem}
\newtheorem{proposition}{Proposition}
\newtheorem{assumption}{Assumption}

\title{Simultaneous estimation of normal means with side information}
\author[1]{Sihai Dave Zhao}
\affil[1]{Department of Statistics, University of Illinois at Urbana-Champaign}

\begin{document}
\maketitle

\begin{abstract}
  The integrative analysis of multiple datasets is an important strategy in data analysis. It is increasingly popular in genomics, which enjoys a wealth of publicly available datasets that can be compared, contrasted, and combined in order to extract novel scientific insights. This paper studies a stylized example of data integration for a classical statistical problem: leveraging side information to estimate a vector of normal means. This task is formulated as a compound decision problem, an oracle integrative decision rule is derived, and a data-driven estimate of this rule based on minimizing an unbiased estimate of its risk is proposed. The data-driven rule is shown to asymptotically achieve the minimum possible risk among all separable decision rules, and it can outperform existing methods in numerical properties. The proposed procedure leads naturally to an integrative high-dimensional classification procedure, which is illustrated by combining data from two independent gene expression profiling studies.
\end{abstract}

\section{\label{sec:intro}Introduction}
Methods for the integrative analysis of multiple datasets are becoming increasingly important. This is especially true in genetics and genomics, where petabytes of public data are readily available for integrative analysis \citep{richardson2016statistical, ritchie2015methods}. For example, \citet{pickrell2016detection} analyzed summary statistics from genome-wide association studies of 42 human traits and found that multiple traits were influenced by several hundred common genetic variants. In a cross-species example, \citet{shpigler2017deep} combined results from a honey bee gene expression study with a database of autism-associated genetic variants and found evidence for evolutionary conservation of genes associated with both honey bee sociality and human autism spectrum disorder. Comparing and contrasting existing data, or combining them with new data, can lead to novel insights that would have been difficult or impossible to uncover with a single dataset alone \citep{tseng2015integrating}.

Integrative analysis strategies can take many forms, and one particularly common implementation is to leverage side information from one or several auxiliary studies for the purpose of improving the analysis of some primary dataset of interest. Examples abound in the multiple testing literature, where methods such as $p$-value weighting and false discovery rate regression incorporate auxiliary information to improve the power to detect true signals in a primary dataset \citep{genovese2006false, ramdas2017unified}. In the genomic risk prediction literature, \citet{hu2017joint} and \citet{zhao2017integrative} showed that summary statistics from previously conducted genome-wide association studies can be used to improve the performance of polygenic risk scores.

Growing interest in these ideas gives rise to an important statistical question: what is the best way to leverage side information? This paper studies this question in a simple but nontrivial problem: the simultaneous estimation of a vector of normal means. The classical version of this problem considers a sequence of independent $X_{i1} \sim N(\theta_{i1}, \sigma_1^2)$ for $i = 1, \ldots, n$ with known $\sigma_1^2$, where the goal is to estimate the $\theta_{i1}$ \citep{johnstone2017gaussian}. The integrative version, studied here, investigates how a auxiliary sequence of Gaussian random variables can be used to improve estimation of the means $\theta_{i1}$ of the primary Gaussian sequence.

This classical Gaussian sequence model is simplistic, but studying data integration in this setting is nevertheless instructive. First, the model is still important for many applications \citep{cai2012minimax, johnstone2017gaussian}. Second, more accurate estimation of the mean vector has immediate implications for high-dimensional classification in genomics \citep{greenshtein2009application}, which will be demonstrated in Section~\ref{sec:data}. Finally, this simple problem can reveal general statistical phenomena that arise in integrative data analysis. More complicated variations of the Gaussian sequence model have been studied, for example involving unknown variances that differ across different indices $i$; see Section \ref{sec:previous}. Extensions of the present work to these more realistic settings are important directions for future work.

Section~\ref{sec:problem} formalizes this integrative estimation task as a compound decision problem and summarizes previous related work. The optimal way to leverage side information is derived in Section~\ref{sec:oracle}, which presents an oracle integrative decision rule that achieves the best risk within a certain class of estimators. This section also introduces a regularized version of the oracle rule that has the same asymptotic risk. A data-driven estimate of this regularized oracle rule is introduced in Section~\ref{sec:data-driven}, and is shown to asymptotically achieve the optimal risk. Its good performance is illustrated in simulations in Section~\ref{sec:sims} and in two genomic risk prediction problems in Section~\ref{sec:data}. A discussion is presented in Section~\ref{sec:disc} and additional simulations and proofs can be found in the Appendix.

\section{\label{sec:problem}Normal means problem with side information}
\subsection{\label{sec:statement}Problem statement}
As in the classical Gaussian sequence problem, consider a sequence of independent $X_{i1} \sim N(\theta_{i1}, \sigma_1^2)$ for $i = 1, \ldots, n$, with $\sigma_1^2$ known. The side information problem studied in this paper further supposes that a second sequence of independent $X_{i2} \sim N(\theta_{i2}, \sigma_2^2), i = 1, \ldots, n$ is available, with $\sigma_2^2$ known. The goal is to estimate the $\theta_{i1}$, just as in the classical problem, but here both the $X_{i1}$ and the $X_{i2}$ can be used for estimation. In this sense, the $X_{i1}$ play the role of a primary dataset, and the $X_{i2}$ act as side information from an auxiliary dataset. This paper assumes that the $X_{i1}$ are independent of the $X_{i2}$ for each $i$, though extensions to dependent $X_{i1}$ and $X_{i2}$ are discussed in Section \ref{sec:disc}.

This formulation is motivated by applications in integrative genomics. The indices $i$ represent different genomic features, such as different genes, and the $X_{i1}$ and $X_{i2}$ represent different measurements on feature $i$ from different studies. For example, in the genomics classification problem described in Section \ref{sec:data}, each $X_{i1}$ estimates a classifier parameter $\theta_{i1}$ corresponding to the $i$th gene from a primary study of interest, and each $X_{i2}$ is the $Z$-score for the $i$th gene reported by an auxiliary study of a related phenotype. The goal is to improve classification accuracy in the primary study by leveraging both $X_{i1}$ and $X_{i2}$ to better estimate the $\theta_{i1}$.

In the above example, the $X_{i1}$ and $X_{i2}$ are paired for each $i$, as both correspond to the same genomic feature. The informativeness of this pairing is crucial for the good performance of data integration. For example, because the phenotypes considered by the two studies in Section \ref{sec:data} are related, genes with significant $Z$-scores in the auxiliary study are also likely to be important features for classification in the primary study, so combining the studies is likely to be fruitful. In contrast, if the phenotypes were unrelated, $X_{i2}$ would likely not be informative about $\theta_{i1}$. The challenge is to develop an estimation procedure that can make optimal use of $X_{i2}$, incorporating them when appropriate and discarding them otherwise. This is addressed by the method proposed in this paper.

To more formally state the problem, define $\bX_{\cdot d} = (X_{1d}, \ldots, X_{nd})$, $\btheta_{\cdot d} = (\theta_{1d}, \ldots, \theta_{nd})$ for $d = 1, 2$, and $\btheta = (\btheta_{\cdot 1}, \btheta_{\cdot 2})$. Then the normal means problem with side information is to find a decision rule $\bdelta(\bX_{\cdot 1}, \bX_{\cdot 2}) = \{\delta_1(\bX_{\cdot 1}, \bX_{\cdot 2}), \ldots, \delta_n(\bX_{\cdot 1}, \bX_{\cdot 2})\} : \mathbb{R}^{2n} \rightarrow \mathbb{R}^n$ that minimizes the risk function
\begin{equation}
  \label{eq:risk}
  R_n(\btheta, \bdelta) = \frac{1}{n} \sum_{i = 1}^n E [\{\theta_{i1} - \delta_i(\bX_{\cdot 1}, \bX_{\cdot 2})\}^2]
\end{equation}
over some class of decision rules. An important class, namely the class of separable estimators, will be considered in this paper and is discussed in Section~\ref{sec:oracle}. This paper adopts the frequentist framework where the $\btheta_{\cdot d}$ are fixed nonrandom constants. The auxiliary data are thus statistically independent of the primary data, and it is interesting that they can still provide useful information for estimating $\btheta_{\cdot 1}$.

To illustrate the complexities of this problem, first suppose that it were known that $\theta_{i2} = \theta_{i1}$ for all $i = 1, \ldots, n$ and that $\sigma_1 = \sigma_2$. The best way to integrate the auxiliary dataset would clearly be to apply existing optimal estimation methods for a single Gaussian sequence to the sequence of averaged observations $(X_{i1} + X_{i2}) / 2$. Next consider a slightly more complicated setting: $\theta_{i2} = \theta_{i1}$ for all but one $i$, but the $i$ for which $\theta_{i2} \ne \theta_{i1}$ is unknown. The auxiliary sequence is clearly still informative for estimating the $\btheta_{\cdot 1}$, but how it should be used is no longer obvious. Finally, consider an even more complicated scenario where $\theta_{i2} = h(\theta_{i1}) + e_i$ for some unknown function $h(t)$, where the $e_i$ are unknown perturbations that exhibit no patterns with respect to $\theta_{i1}$. If the magnitudes of the $e_i$ are small relative to the $\theta_{i1}$, $\bX_{\cdot 2}$ should still be useful when estimating $\btheta_{\cdot 1}$, but it is even less clear how to optimally integrate it into the estimation procedure. This paper provides one approach.

\subsection{\label{sec:previous}Previous work}
The classical normal means estimation problem without side information, which aims to minimize the risk function \eqref{eq:risk} using decision rules that can depend only on $\bX_{\cdot 1}$ and not $\bX_{\cdot 2}$, has inspired an enormous literature \citep{johnstone2017gaussian}. \citet{stein1956inadmissibility} found that the maximum likelihood estimator $\delta_i(\bX_{\cdot 1}) = X_{i1}$ is inadmissible, and since then research has focused on finding alternative estimators with better risk properties. Several different but intimately related perspectives on this problem have been developed.

The shrinkage perspective is exemplified by the James-Stein estimator \citep{james1961estimation, stigler19901988}, which estimates $\theta_{i1}$ by scaling $X_{i1}$ towards zero. The empirical Bayes perspective \citep{robbins1985empirical} treats the $\theta_{i1}$ as random draws from a prior distribution, uses the $X_{i1}$ to estimate any unknown parameters in the prior, then estimates each $\theta_{i1}$ by its posterior mean conditional on $X_{i1}$. \citet{efron1973stein} showed that the James-Stein estimator is an empirical Bayes estimator assuming a normal prior for the $\theta_{i1}$. The compound decision perspective \citep{robbins1951asymptotically, zhang1997empirical} treats the $\theta_{i1}$ as nonrandom constants and directly derives the decision rule that minimizes the risk. Under certain conditions, the optimal solution from this perspective is closely related to nonparametric empirical Bayes estimators \citep{brown2009nonparametric, jiang2009general, zhang2003compound}.

More complicated versions of the classical normal means problem have also been intensely studied. For example, specialized methods have been developed for estimating sparse normal means, where most of the $\theta_{i1}$ are assumed to equal zero \citep{castillo2012needles, donoho1994ideal, donoho1995adapting, martin2014asymptotically}. Heteroscedastic normal sequences, where the $X_{i1}$ can have different variances for different indices $i$, have also been considered, both when the variances are known \citep{fu2019nonparametric, tan2016steinized, weinstein2018group, xie2012sure, zhang2017empirical} and when they are unknown but estimates are available \citep{feng2018approximate, gu2017empirical, jing2016sure}.

So far, however, most work on the normal means problem and its variants has considered only a single sequence of observations $X_{i1}$, and it appears that the side information problem \eqref{eq:risk} has not yet been widely studied. \citet{jiang2010empirical}, \citet{cohen2013empirical}, \citet{tan2016steinized}, and \citet{kou2017optimal} proposed methods that can integrate $X_{i2}$, but these essentially require knowledge of the nature of the relationship between $\theta_{i1}$ and $X_{i2}$, and may not work well when this relationship is misspecified. \citet{banerjee2018adaptive} studied the side information problem, but only for sparse $\btheta_{\cdot 1}$. Very recently \citet{saha2017nonparametric} and \citet{koudstaal2018multiple} considered two or more Gaussian sequences, but minimized the risk of estimating the means of all of the sequences, rather than the means of just one of them as in \eqref{eq:risk}. In contrast to existing work, this paper studies the optimal use of $\bX_{\cdot 1}$ and $\bX_{\cdot 2}$ for estimating possibly non-sparse $\btheta_{\cdot 1}$.

\section{\label{sec:oracle}Oracle integrative separable rules}
Without any restrictions, the optimal decision rule is simply $\delta_i(\bX_{\cdot 1}, \bX_{\cdot 2}) = \theta_{i1}$, which is not useful because the performance of this rule cannot realistically be achieved using the observed data alone. Instead, this paper only considers rules in the class
\begin{equation}
  \label{eq:S}
  \mathcal{S} = \{ \bdelta(\bX_{\cdot 1}, \bX_{\cdot 2}) : \delta_i(\bX_{\cdot 1}, \bX_{\cdot 2}) = f(X_{i1}, X_{i2})\},
\end{equation}
where $f$ is some fixed real-valued function that is applied to each pair $(X_{i1}, X_{i2})$ in order to estimate $\theta_{i1}$. In other words, the estimate of $\theta_{i1}$ is calculated by applying $f(x_1, x_2)$ to only the $i$th pair of observations $(X_{i1}, X_{i2})$, and $f(x_1, x_2)$ cannot vary with $i$.

Rules in $\mathcal{S}$, called ``separable'' rules, are appealing because of their simplicity and have been extensively studied \citep{brown2009nonparametric, cai2012minimax, robbins1951asymptotically, zhang2003compound}. The maximum likelihood estimator $\bdelta(\bX_{\cdot 1}, \bX_{\cdot 2}) = X_{i1}$ belongs to $\mathcal{S}$, and the James-Stein estimator approximates the optimal separable rule that is linear in $X_{i1}$ \citep{jiang2009general}. The minimum risk among all separable estimators has been shown to be asymptotically equivalent, in a certain sense, to the minimum achievable risk over the larger class of permutation invariant estimators \citep{greenshtein2009asymptotic}.

The following proposition describes the oracle optimal integrative rule in $\mathcal{S}$ for estimating $\btheta_{\cdot 1}$, which will perform no worse than any separable rule that relies only on $\bX_{\cdot 1}$. It is a direct consequence of the fundamental theorem of compound decision problems \citep{robbins1951asymptotically, jiang2009general}. Let $\phi(x)$ denote the standard normal density and define
\begin{equation}
  \label{eq:density}
  \begin{aligned}
    p(x_1, x_2; t_1, t_2)
    =\,&
    \frac{1}{\sigma_1} \phi \left( \frac{x_1 - t_1}{\sigma_1} \right)
    \frac{1}{\sigma_2} \phi \left( \frac{x_2 - t_2}{\sigma_2} \right),\\
    p_i^0(x_1, x_2)
    =\,&
    p(x_1, x_2; \theta_{i1}, \theta_{i2}),
  \end{aligned}
\end{equation}
so that the density of $(X_{i1}, X_{i2})$ can be abbreviated by $p_i^0(x_1, x_2)$. As mentioned in the problem statement in Section \ref{sec:statement}, this paper assumes that the $X_{i1}$ and $X_{i2}$ are independent, but the following result is easily extended to settings where $X_{i1}$ and $X_{i2}$ are correlated; see Section \ref{sec:disc}.

\begin{proposition}
  \label{prop:oracle}
  Define the decision rule $\bdelta^\star = (\delta^\star_1, \ldots, \delta^\star_n)$ where $\delta^\star_i(\bX_{\cdot 1}, \bX_{\cdot 2}) = f^\star(X_{i1}, X_{i2})$ and
  \begin{equation}
    \label{eq:oracle}
    f^\star(x_1, x_2)
    =
    \frac{
      \sum_{j = 1}^n \theta_{j1} p_j^0(x_1, x_2)
    }{
      \sum_{j = 1}^n p_j^0(x_1, x_2)
    }.
  \end{equation}
  Then $R_n(\btheta, \bdelta) \geq R_n(\btheta, \bdelta^\star)$ for any $\bdelta \in \mathcal{S}$ \eqref{eq:S} for $R_n(\btheta, \bdelta)$ in \eqref{eq:risk}.
\end{proposition}

The oracle rule $\bdelta^\star$ also has a useful interpretation as a Bayes rule. If the $\theta_{i1}$ are viewed as independent draws from the discrete prior distribution
\begin{equation}
  \label{eq:G}
  G_n(t_1, t_2) = \frac{1}{n} \sum_{i = 1}^n I(\theta_{i1} \leq t_1, \theta_{i2} \leq t_2),
\end{equation}
then the posterior expectation $E(\theta_{i1} \mid X_{i1}, X_{i2})$ of $\theta_{i1}$ is exactly equal to \eqref{eq:oracle}. This is an example of the close connection between compound decision problems and nonparametric empirical Bayes procedures. The dependence between $\theta_{i1}$ and $\theta_{i2}$ under $G_n$ quantifies the amount of information that can be borrowed from $X_{i2}$.

While appealing, this Bayesian interpretation is not necessary for Proposition \ref{prop:oracle}, which holds for fixed and constant $\btheta_{\cdot 1}$ and $\btheta_{\cdot 2}$. Interestingly, under this frequentist setting Proposition \ref{prop:oracle} shows that $\bX_{\cdot 2}$ can improve the estimation of $\btheta_{\cdot 1}$ even though $\bX_{\cdot 1}$ and $\bX_{\cdot 2}$ are statistically independent, as long the sequences $\btheta_{\cdot 1}$ and $\btheta_{\cdot 2}$ are related in some sense. There need not be an obvious functional relationship between the two mean vectors.

The above view of side information is slightly different from that of existing frameworks. Previous methods \citep{jiang2010empirical, kou2017optimal, tan2016steinized} posit some functional relationship, typically linear, between $\theta_{i1}$ and the observed $X_{i2}$, rather than between $\theta_{i1}$ and the true mean $\theta_{i2}$. For example, \citet{kou2017optimal} assume that $\theta_{i1} = h(X_{i2}) + e_i$ for some error term $e_i$, where $h(x)$ must be known up to a finite-dimensional parameter. These methods treat the $X_{i2}$ as fixed, while the proposed framework acknowledges that the $X_{i2}$ are random variables. The difference between existing work and the present setting is akin to the difference between classical regression methods and those that take into account covariate measurement error.

\begin{figure}[h!]
  \centering
  \includegraphics[width = \textwidth]{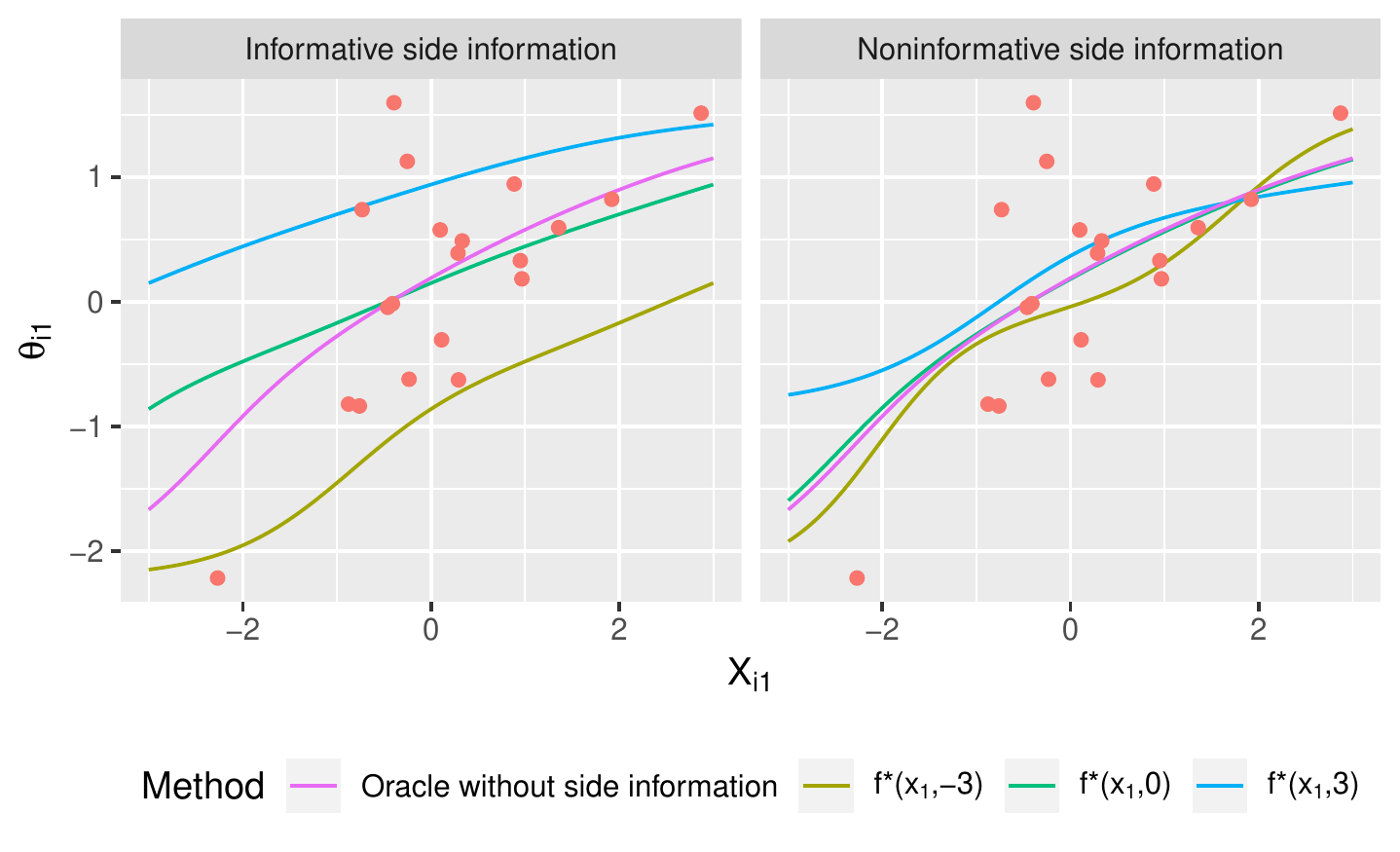}\par
  \caption{\label{fig:estimators}Oracle estimators with and without side information for $n = 20$ pairs $(X_{i1}, X_{i2})$. Each curve plots the estimate of $\theta_{i1}$ as a function of $X_{i1}$. Each dot corresponds to an observed $X_{i1}$ along with its true mean $\theta_{i1}$.}
\end{figure}

Figure \ref{fig:estimators} illustrates the oracle rule $\bdelta^\star$ \eqref{eq:oracle} and compares it to the best separable estimator that does not use $\bX_{\cdot 2}$, which is the posterior expectation of $\theta_{i1}$ under the prior $G_n$ conditional only on $X_{i1}$ \citep{zhang2003compound}. In both panels, $\theta_{i1}$ was generated by drawing $n = 20$ values from a standard normal distribution. In the left panel, $\btheta_{\cdot 2} = \btheta_{\cdot 1}$, so $\bX_{\cdot 2}$ was highly informative for $\btheta_{\cdot 1}$. Thus $f^\star(x_1, 3)$ gave the best estimates of $\theta_{i1}$ for large $X_{i1}$ and $f^\star(x_1, -3)$ was most accurate for small $X_{i1}$. In the right panel of Figure \ref{fig:estimators}, the $\theta_{i2}$ were generated from an independent standard normal so that $\bX_{\cdot 2}$ was completely uninformative. In this non-informative setting, $\bdelta^\star$ may not have the same performance as the optimal non-integrative separable rule for any given set of $\bX_{\cdot 1}$ and $\bX_{\cdot 2}$, but in expectation Proposition \ref{prop:oracle} guarantees that it will have equal or lower risk.

The oracle separable integrative rule $\bdelta^\star$ described in \eqref{eq:oracle} cannot be implemented in practice because it requires knowing the true $(\theta_{i1}, \theta_{i2})$ up to permutation of the indices. Section~\ref{sec:data-driven} will introduce a data-driven rule that targets the performance of $\bdelta^\star$, though for technical reasons it will be more convenient to target a regularized version of the oracle rule. This will be denoted by $\bdelta^\star_\rho = (\delta^\star_{\rho 1}, \ldots, \delta^\star_{\rho n})$, with $\delta^\star_{\rho i}(\bX_{\cdot 1}, \bX_{\cdot 2}) = f^\star_\rho(X_{i1}, X_{i2})$ for
\begin{equation}
  \label{eq:regularized}
  f^\star_\rho(x_1, x_2)
  =
  x_1
  +
  \frac{
    \sum_{j = 1}^n (\theta_{j1} - x_1) p_j^0(x_1, x_2)
  }{
    \rho + \sum_{j = 1}^n p_j^0(x_1, x_2)
  }
\end{equation}
and $\rho$ a small positive constant that prevents the denominator from being too close to zero. Under some assumptions, $\bdelta^\star_\rho$ will have the same asymptotic risk as the oracle $\bdelta^\star$.

\begin{assumption}
  \label{a:theta}
  There exist positive constants $C$ and $\eta$ such that $\vert \theta_{id} \vert \leq C n^{1/4 - \eta}$ for $i = 1, \ldots, n$ and $d = 1, 2$.
\end{assumption}

\begin{theorem}
  \label{thm:regularized}
  Under Assumption \ref{a:theta}, $\lim_{n \rightarrow \infty} \{ R_n(\btheta, \bdelta^\star_\rho) - R_n(\btheta, \bdelta^\star) \} = 0$.
\end{theorem}

Assumption \ref{a:theta} determines how quickly the magnitudes of $\theta_{id}$ can grow. To put this rate into perspective, if the $\theta_{id}$ were random draws from a normal distribution, then $\max_i \vert \theta_{id} \vert$ would be $O(\log^{1/2} n)$ almost surely. Related assumptions, which essentially restrict how variable the $\theta_{id}$ can be, have been made in previous work on normal means estimation without side information. For example, \citet{xie2012sure} require $\lim n^{-1} \sum_i \theta_{i1}^2 < \infty$, and \citet{jiang2009general} and \citet{zhang2009generalized} control the rate of the $p$-th weak moment of the distribution function $n^{-1} \sum_i I(\theta_{i1} \leq t_1)$.

\section{\label{sec:data-driven}Data-driven separable estimator}
\subsection{\label{sec:npeb}Existing nonparametric empirical Bayes approach}
By Proposition \ref{prop:oracle} and Theorem \ref{thm:regularized}, the regularized oracle $\bdelta^\star_\rho$ \eqref{eq:regularized} is asymptotically optimal within the class of separable estimators \eqref{eq:S}, but it cannot be implemented in practice. It therefore remains to develop a fully data-driven estimator for the $\theta_{i1}$. Two classes of approaches already exist. They have been termed $f$- and $g$-modeling \citep{efron2014two, efron2019bayes} and are based on nonparametric empirical Bayes principles that pretend that the $(\theta_{i1}, \theta_{i2})$ are random variables with prior distribution $G_n(t_1, t_2)$ \eqref{eq:G}.

In $f$-modeling, the oracle estimator \eqref{eq:oracle} would be re-expressed as
\[
f^\star(x_1, x_2)
=
x_1 + \frac{\tilde{p}'(x_1, x_2)}{\tilde{p}(x_1, x_2)},
\]
where $\tilde{p}'(x_1, x_2) = \partial \tilde{p} / \partial x_1$ and
\[
\tilde{p}(x_1, x_2) = \int p(x_1, x_2; t_1, t_2) dG_n(t_1, t_2)
\]
with $p(x_1, x_2; t_1, t_2)$ from \eqref{eq:density}. If the $(\theta_{i1}, \theta_{i2})$ were truly random, $\tilde{p}(x_1, x_2)$ could be interpreted as the marginal density of $(X_{i1}, X_{i2})$, and $\tilde{p}(x_1, x_2)$ and $\tilde{p}'(x_1, x_2)$ could be estimated nonparametrically using kernel density estimators. In $g$-modeling, the oracle estimator would be re-expressed as
\[
f^\star(x_1, x_2) = x_1 + \frac{\int (t_1 - x_1) p(x_1, x_2; t_1, t_2) dG_n(t_1, t_2)}{\int p(x_1, x_2; t_1, t_2) dG_n(t_1, t_2)},
\]
and if the $(\theta_{i1}, \theta_{i2})$ were truly random, a nonparametric estimate of $G_n(t_1, t_2)$ could be obtained by maximizing the marginal log-likelihood \citep{kiefer1956consistency}
\[
\argmax_G \prod_{i = 1}^n \int p(X_{i1}, X_{i2}; t_1, t_2) dG(t_1, t_2).
\]

Both $f$- and $g$-modeling have been used in normal means problems without side information, where they are asymptotically optimal even in the frequentist framework where the $\theta_{i1}$ and $\theta_{i2}$ are nonrandom \citep{brown2009nonparametric, feng2018approximate, fu2019nonparametric, jiang2009general, koenker2014gaussian, koenker2014convex, saha2017nonparametric, zhang2009generalized}. However, neither approach directly estimates the oracle decision rule, with $f$-modeling proceeding through the intermediate quantity $\tilde{p}(x_1, x_2)$ and $g$-modeling proceeding through $G_n(t_1, t_2)$.

\subsection{\label{sec:proposed}Proposed direct risk minimization approach}
This paper explores a more direct approach to estimating the oracle integrative separable classifier. Motivated by the regularized oracle \eqref{eq:regularized}, consider separable rules of the form $\bdelta^t_\rho = (\delta^t_{\rho 1}, \ldots, \delta^t_{\rho n})$ with
\begin{equation}
  \label{eq:delta^t}
  \delta^{\bt}_{\rho i}(x_1, x_2)
  =
  x_1
  +
  \frac{
    \sum_{j = 1}^n (t_{j1} - x_1) p(x_1, x_2; t_{j1}, t_{j2})
  }{
    \rho + \sum_{j = 1}^n p(x_1, x_2; t_{j1}, t_{j2})
  },
\end{equation}
for a given $\bt = (t_{11}, \ldots, t_{n1}, t_{12}, \ldots, t_{n2})$. By Theorem \ref{thm:regularized}, the optimal $\bt$ equals $(\theta_{11}, \ldots, \theta_{n1}, \theta_{12}, \ldots, \theta_{n2})$, but the $\theta_{id}$ are not known. The challenge is to choose a $\bt$ in a data-driven fashion that can still asymptotically achieve the optimal risk.

Choosing $\bt$ to minimize the risk \eqref{eq:risk} of $\bdelta^{\bt}_\rho$ \eqref{eq:delta^t} should give an estimator with good performance. However, calculating the risk requires knowing the true $\theta_{id}$. On the other hand, Stein's Lemma \citep{stein1981estimation} can be used to give an unbiased estimate of the true risk as a function only of $\bt$:
\begin{equation}
  \label{eq:sure}
  \begin{aligned} 
    & \textsc{sure}(\bt)\\
    =\,&
    \frac{2}{n} \sum_{i = 1}^n
    \frac{
      \sum_j (t_{j1} - X_{i1})^2 p(X_{i1}, X_{i2}; t_{j1}, t_{j2})
    }{
      \rho + \sum_j p(X_{i1}, X_{i2}; t_{j1}, t_{j2})
    }
    -
    \frac{2}{n} \sigma_1^2 \sum_{i = 1}^n
    \frac{
      \sum_j p(X_{i1}, X_{i2}; t_{j1}, t_{j2})
    }{
      \rho + \sum_j p(X_{i1}, X_{i2}; t_{j1}, t_{j2})
    }
    \,- \\
    &
    \frac{1}{n} \sum_{i = 1}^n
    \left\{
    \frac{
      \sum_j (t_{1j} - X_{i1}) p(X_{i1}, X_{i2}; t_{j1}, t_{j2})
    }{
      \rho + \sum_j p(X_{i1}, X_{i2}; t_{j1}, t_{j2})
    }
    \right\}^2
    +
    \sigma_1^2.
  \end{aligned}
\end{equation}
The following theorem shows that $\textsc{sure}(\bt)$ is also a good approximation to the actual loss
\begin{equation}
  \label{eq:loss}
  \ell_n(\bt) = \frac{1}{n} \sum_{i = 1}^n \{\theta_{i1} - \delta^{\bt}_{\rho i}(X_{i1}, X_{i2})\}^2
\end{equation}
uniformly over the set
\begin{equation}
  \label{eq:T}
  \mathcal{T} = \{ \bt : \vert t_{jd} \vert \leq C n^{1/4 - \eta}, j = 1, \ldots, n, d = 1, 2 \}.
\end{equation}
\begin{theorem}
  \label{thm:uniform}
  Under Assumption \ref{a:theta}, if $0 < \rho \leq 1$, then
  \[
  \lim_{n \rightarrow \infty} E \sup_{\bt \in \mathcal{T}} \vert \textsc{sure}(\bt) - \ell_n(\bt) \vert = 0.
  \]
\end{theorem}

The tuning parameter $\bt$ can now be chosen by minimizing this estimated risk, as a proxy for minimizing the unknown true risk. The proposed estimator is therefore defined to be
\begin{equation}
  \label{eq:proposed}
  \bdelta^{\hat{\bt}}_\rho \mbox{ as in \eqref{eq:delta^t} with } \hat{\bt} = \argmin_{\bt \in \mathcal{T}} \textsc{sure}(\bt).
\end{equation}
This strategy of direct risk minimization is common in the compound decision literature \citep{jing2016sure, kou2017optimal, tan2016steinized, weinstein2018group, xie2012sure, xie2016optimal, zhang2017empirical}, but has not yet been used to approximate an optimal separable rule like \eqref{eq:regularized}. The following theorem shows that \eqref{eq:proposed} can asymptotically achieve the same the performance as the optimal separable decision rule.

\begin{theorem}
  \label{thm:opt_risk}
  Under the same conditions as Theorem \ref{thm:uniform}, $\lim_{n \rightarrow \infty} \{ E \ell_n(\hat{\bt}) - R_n(\btheta, \bdelta^\star) \} \leq 0$, where $E \ell_n(\hat{\bt})$ is the risk of the proposed estimator $\bdelta^{\hat{\bt}}_\rho$ \eqref{eq:proposed}.
\end{theorem}

\subsection{\label{sec:implementation}Implementation}
The proposed estimator has been implemented in the R package \verb|cole|, available at \verb|github.com/sdzhao/cole|. In practice, the exact value of $\rho$ appears to make little difference, and $\rho = 0$ works well in most cases. When the range of the $X_{id}$ is very large or the variances $\sigma_d^2$ are very small, problem may arise when calculating $\textsc{sure}(\bt)$ due to numerical precision, in which case setting $\rho = 10^{-12}$ seems to be sufficient. Throughout this paper, the proposed method was implemented with $\rho = 0$.

Because the value of $C n^{1/4 - \eta}$ that defines the feasible set $\mathcal{T}$ \eqref{eq:T} is not known, in practice the minimization in \eqref{eq:proposed} can be performed over
\[
\hat{\mathcal{T}}
=
\prod_{i = 1}^n
[X_{i1} - M \sigma_1, X_{i1} + M \sigma_1] \times [X_{i2} - M \sigma_2, X_{i2} + M \sigma_2]
\]
for some sufficiently large positive constant $M$, so that $\hat{\mathcal{T}}$ contains $(\btheta_{\cdot 1}, \btheta_{\cdot 2})$ with probability $\Phi(-M)^n$, where $\Phi$ is the cumulative distribution function of a standard normal. By default, \verb|cole| uses $M = 5$, so that $\hat{\mathcal{T}}$ contains $(\btheta_{\cdot 1}, \btheta_{\cdot 2})$ with probability 0.99 when $n = 10,000$. Optimizing $\textsc{sure}(\bt)$ over $\hat{\mathcal{T}}$ is sensible because it is known from Theorem \ref{thm:regularized} that $E\{\textsc{sure}(\bt)\}$ achieves a global minimum at $t_{jd} = \theta_{jd}$. This method works well, but bridging the gap between the theoretical procedure and its practical implementation is an important direction for future work.

Minimizing $\textsc{sure}(\bt)$ is difficult because it is a nonconvex function. The implementation in \verb|cole| performs a simple coordinate descent. At initialization, $t_{id}$ is set to $X_{id}$, and at each iteration one $t_{id}$ is updated by optimizing over $K$ equally spaced candidates in $[X_{id} - M \sigma_d, X_{id} + M \sigma_d]$. By default, \verb|cole| uses $K = 10$, and all analyses in this paper use $K = 10$ unless otherwise stated. The coordinates of $\bt$ are updated in the order $t_{11}, t_{21}, \ldots, t_{n1}, t_{12}, \ldots, t_{n2}$, and convergence is reached when all of the coordinates have been cycled through once without changing the value of $\textsc{sure}(\bt)$ by more than a small  $\epsilon$, which \verb|cole| sets to $10^{-5}$ by default.

\section{\label{sec:sims}Simulations}
\subsection{\label{sec:sims_classical}Normal means problem without side information}
The direct risk minimization approach proposed in this paper for estimating optimal separable decision rules appears to be novel in the compound decision literature. This section thus first illustrates how this idea performs in the classical normal means problem without side information. The optimal separable estimator and its corresponding unbiased risk estimate will look like \eqref{eq:oracle} and \eqref{eq:sure}, respectively, but with the density $p(x_1, x_2; t_1, t_2)$ replaced by $\phi\{(x_1 - t_1) / \sigma_1\} / \sigma_1$, where $\phi(x)$ is the standard normal density. Similar to \eqref{eq:proposed}, a data-driven estimator of the oracle rule can be obtained by minimizing the risk estimate over $\bt_1$ using the coordinate descent algorithm described in Section \ref{sec:implementation}; this is available in the \verb|cole| package. Analogs of Theorems \ref{thm:regularized}--\ref{thm:opt_risk} can also be proved.

The direct estimator was compared to the $g$-modeling procedure of \citet{jiang2009general}, which is also asymptotically risk-optimal. One independent sequence $X_{i1}, i = 1, \ldots, 1,000$ was generated from N$(\theta_{i1}, 1)$, with the goal of estimating $\btheta_{\cdot 1}$ using only $\bX_{\cdot 1}$. The $\theta_{i1}$ equaled either 0 or $\mu$ and the number of nonzero $\theta_{i1}$ equaled either 5, 50, or 500. Table \ref{tab:nonintegrative} displays the average total squared errors over 100 replications. Results for the estimator of \citet{jiang2009general} were taken directly from their Table 1, while the proposed estimator was implemented using a coordinate descent algorithm that optimized over $K = 50$ candidates for each $t_{1j}$. The results show that both estimators had almost identical performance.

\begin{table}[ht]
  \centering
  \begin{tabular}{l|cccc|cccc|cccc}
  \hline
  \hline
  \# nonzero & \multicolumn{4}{c}{5} & \multicolumn{4}{c}{50} & \multicolumn{4}{c}{500} \\
  $\mu$ & 3 & 4 & 5 & 7 & 3 & 4 & 5 & 7 & 3 & 4 & 5 & 7  \\
  \hline
  \hline
  GMLEB & 39 & 34 & 23 & 11 & 157 & 105 & 58 & 14 & 459 & 285 & 139 & 18\\
  Proposed & 37 & 32 & 21 & 11 & 158 & 110 & 56 & 14 & 460 & 289 & 133 & 21 \\ 
  \hline
  \end{tabular}
  \caption{\label{tab:nonintegrative}Average total squared errors for the classical normal means problem without side information. GMLEB: the $g$-modeling method of \citet{jiang2009general}.}
\end{table}

\subsection{Settings for normal means problem with side information}
The primary data $\bX_{\cdot 1} = (X_{11}, \ldots, X_{n1})$ were generated in four different ways, for three dense and one sparse configuration of their means $\btheta_{\cdot 1}$. To generate dense $\btheta_{\cdot 1}$, values of $\theta_{i1}$ were independently drawn from either a N$(0,1)$, a Unif$(-2, 2)$, or an Exp$(1)$ distribution. To generate the sparse configuration, 10\% of the coordinates of $\btheta_{\cdot 1}$ were set to 1.5 and the rest were set to 0. The observed primary data were generated as $X_{i1} = \theta_{i1} + \epsilon_{i1}$, where the $\epsilon_{i1}$ were independently drawn from standard normal random variables. The $\theta_{i1}$ were fixed across all replications.

For each of these four settings, the auxiliary data $\bX_{\cdot 2} = (X_{12}, \ldots, X_{n2})$ were generated in three different ways, to model different degrees of informativeness of $\btheta_{\cdot 2}$. First define $e_i$ to be independent draws from a Unif$(-4, 4)$. To generate strongly, weakly, and non-informative side information, $\theta_{i2}$ was set to be either $2 \theta_{i1}^2$, $\theta_{i1}^2 + e_i$, or $e_i$, respectively. The observed auxiliary data were generated as $X_{i2} = \theta_{i2} + \epsilon_{i2}$, where the $\epsilon_i$ were again independently drawn from standard normal random variables. The $\theta_{i2}$ were fixed across all replications.

The proposed integrative normal means estimator \eqref{eq:proposed} was compared to two existing approaches that can incorporate side information. One was the procedure of \citet{banerjee2018adaptive}. The other was estimator (1) of \citet{kou2017optimal}, defined as
\[
\frac{\lambda}{\lambda + \sigma_1^2} X_{i1} + \frac{\sigma_1^2}{\lambda + \sigma_1^2} h(X_{i2})
\]
for some function $h(x)$ known up to a finite number of parameters. These unknown parameters, as well as $\lambda$, are chosen by minimizing an unbiased estimate of the risk of this estimator. This estimator is motivated by the regression model $\theta_{i1} = h(X_{i2}) + e_i$ for some error terms $e_i$. However, it can be difficult to choose the correct regression function $h(x)$. For example, in some of the present simulation settings, the true relationship between the primary and auxiliary data is $\theta_{i2} = 2 \theta_{i1}^2 + e_i$, which is difficult to translate into a regression model of $\theta_{i1}$ on $X_{i2}$. When implementing the method of \citet{kou2017optimal}, these simulations used both the nonlinear model $\theta_{i1} = \beta_0 + \beta_1 \vert X_{i2} \vert^{1/2} + e_i$ and the linear model $\theta_{i1} = \beta_0 + \beta_1 X_{i2} + e_i$.

Finally, two additional estimators for $\btheta_{\cdot 1}$ were also implemented to provide performance baselines. The first was the oracle \eqref{eq:oracle}, which attains the lowest possible risk of any separable decision rule that incorporates side information. The second was the $g$-modeling method of \citet{jiang2009general}, which can asymptotically achieve the optimal risk of any separable rule that does not use side information.

\subsection{Results for normal means problem with side information}
\begin{figure}[h!]
  \centering
  
  \includegraphics[width = \textwidth]{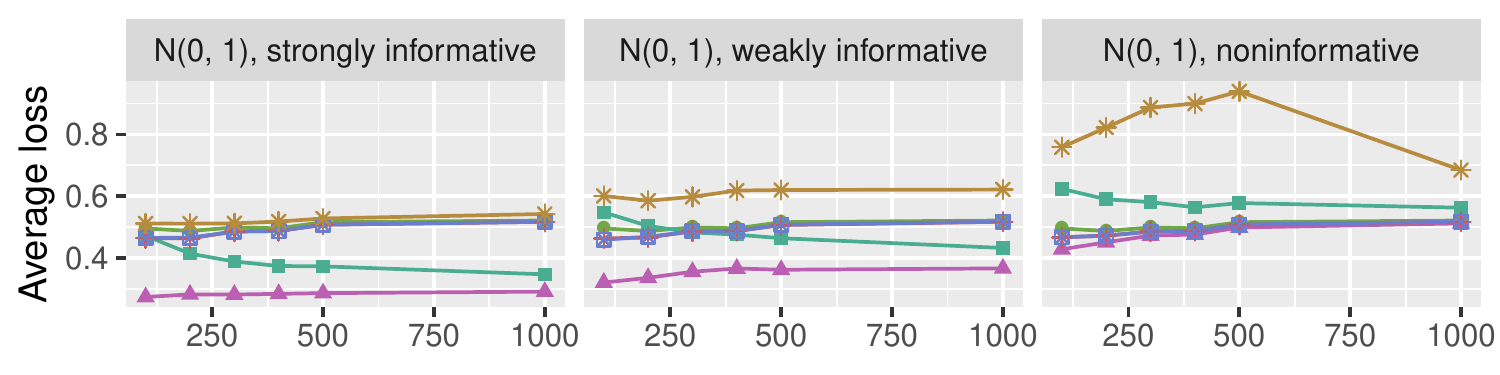}
  
  \includegraphics[width = \textwidth]{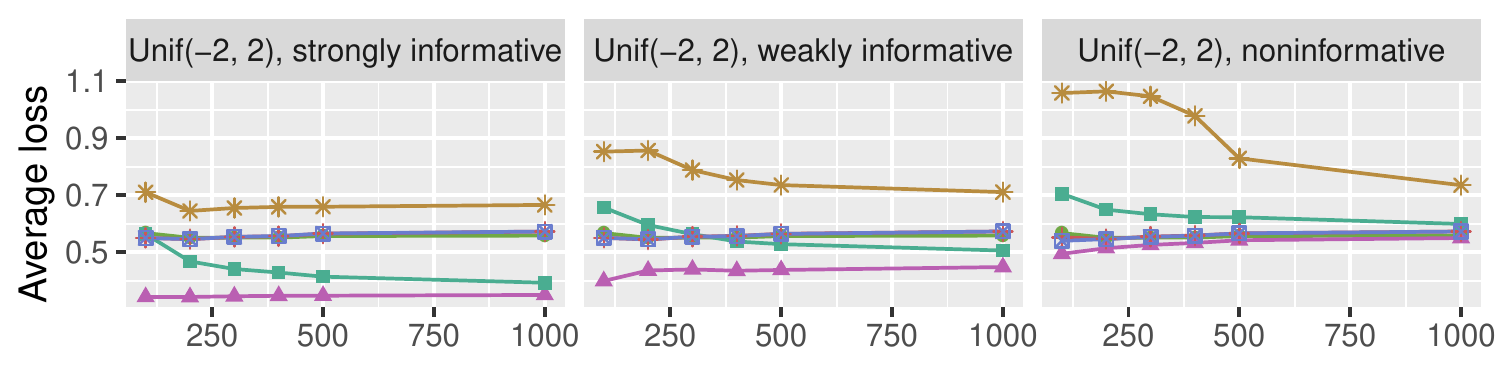}
  
  \includegraphics[width = \textwidth]{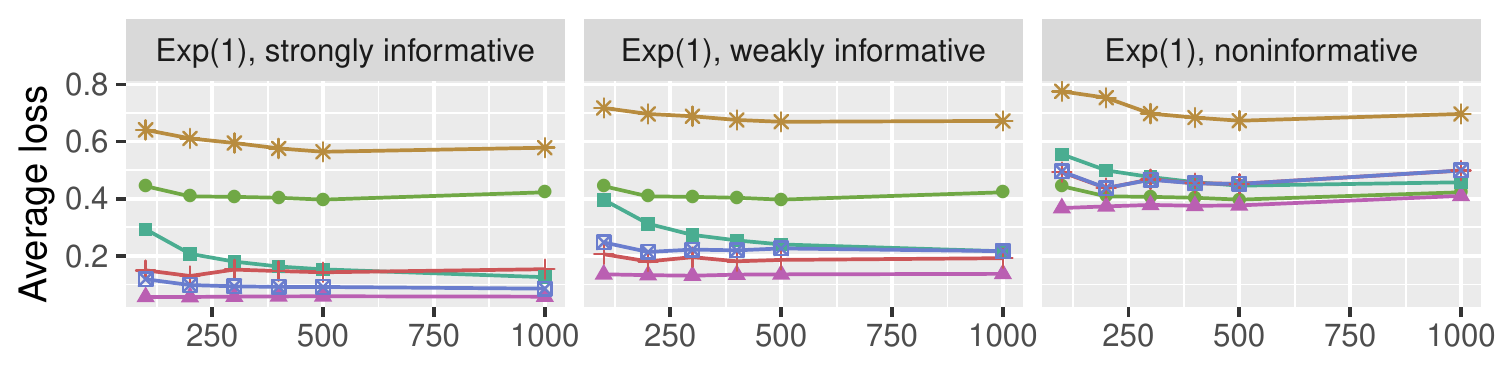}
  
  \includegraphics[width = \textwidth, trim = 0 20 0 0]{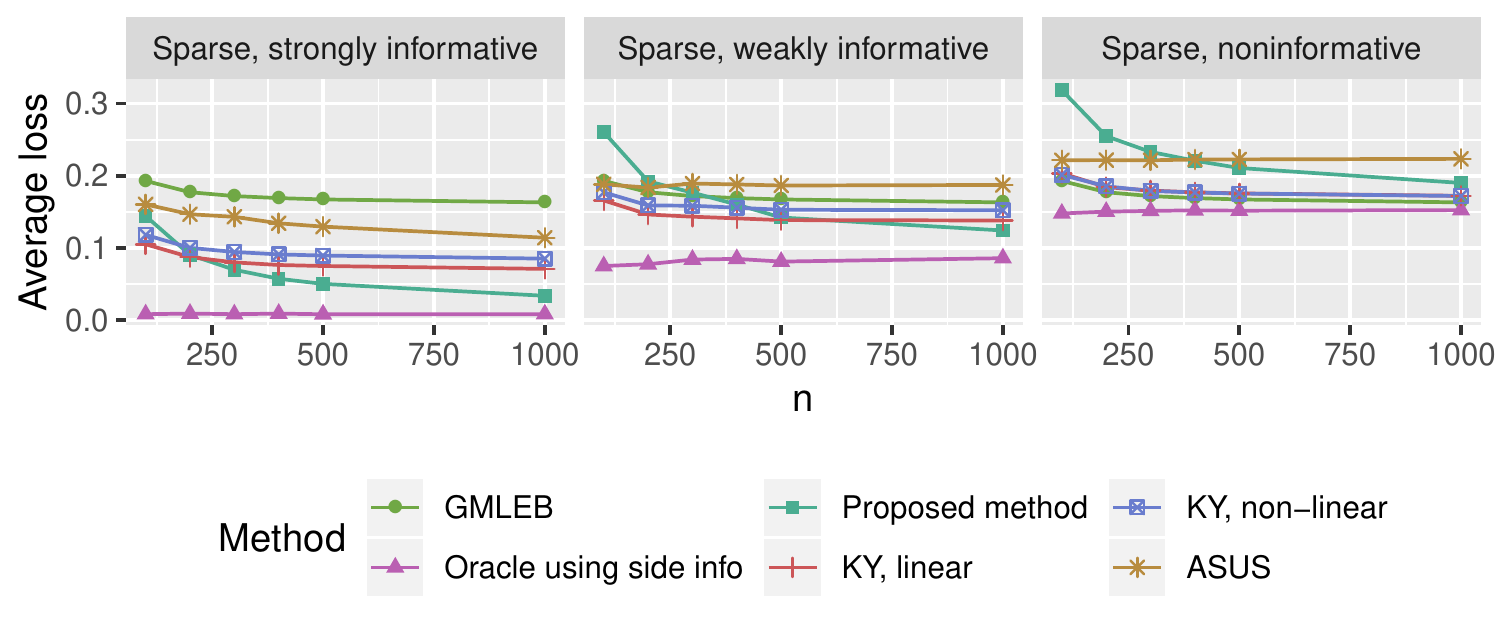} \par
  \caption{\label{fig:sims}Average losses for four different configurations of $\btheta_{\cdot 1}$ and three degrees of informativeness of $\btheta_{\cdot 2}$. GMLEB: method of \citet{jiang2009general}; KY, linear: method of \citet{kou2017optimal} with model $\theta_{i1} = \beta_0 + \beta_1 X_{i2} + e_i$; KY, nonlinear: method of \citet{kou2017optimal} with model $\theta_{i1} = \beta_0 + \beta_1 \vert X_{i2} \vert^{1/2} + e_i$; ASUS: method of \citet{banerjee2018adaptive}.}
\end{figure}

Figure \ref{fig:sims} illustrates the average losses, over 200 simulations, achieved by the competing methods for N$(0,1)$, Unif$(-2, 2)$, Exp$(1)$, or sparse $\btheta_{\cdot 1}$ and non-informative, weakly informative, or strongly informative $\btheta_{\cdot 2}$. Comparing the performances of the oracle rule \eqref{eq:oracle} and the method of \citet{jiang2009general} shows that including auxiliary data does not degrade estimation accuracy asymptotically when $\btheta_{\cdot 2}$ is non-informative, and can greatly improve it when $\btheta_{\cdot 2}$ is informative.

The performance of the proposed data-driven estimator $\bdelta^{\hat{\bt}}_\rho$ \eqref{eq:proposed} indeed appeared to converge to the oracle performance as the number of observations $n$ increased, consistent with Theorem \ref{thm:opt_risk}. Unlike the oracle, however, incorporating non-informative $\bX_{\cdot 2}$ in $\bdelta^{\hat{\bt}}_\rho$ resulted in worse performance compared to the other methods when $n$ was small. This is expected, as non-informative $X_{i2}$ add extra noise without decreasing bias, and the data-driven method requires enough samples to learn that the $X_{i2}$ are not useful. In contrast, $\bdelta^{\hat{\bt}}_\rho$ regained its competitiveness for larger $n$, and when the auxiliary $\bX_{\cdot 2}$ were at least weakly informative, it frequently achieved the lowest risk among all methods. These results suggest that incorporating $\bX_{\cdot 2}$ using the proposed method is highly effective when $\bX_{\cdot 2}$ is informative, and does not do too much harm when it is not.

The proposed $\bdelta^{\hat{\bt}}_\rho$ was sometimes outperformed by the two different implementations of the procedure of \citet{kou2017optimal}, for example when the $\theta_{i1}$ were generated from Exp$(1)$. This may be because this setting was particularly difficult for the proposed method. Out of the four configurations of $\btheta_{\cdot 1}$, the maximum value of $\vert \theta_{i1} \vert$ was largest under the Exp$(1)$ configuration, and Assumption \ref{a:theta} makes it clear that restricting this maximum value is important for the good performance of $\bdelta^{\hat{\bt}}_\rho$. On the other hand, when $n = 1,000$ $\bdelta^{\hat{\bt}}_\rho$ had essentially the same risk as the methods of \citet{kou2017optimal}, and for other configurations of $\theta_{i1}$, $\bdelta^{\hat{\bt}}_\rho$ could perform significantly better.

Finally, the proposed rule performed extremely well with sparse $\btheta_{\cdot 1}$, even though it was not designed for this scenario. When the auxiliary data were strongly informative, it achieved the lowest risks among all implemented methods when $n \geq 200$. It would be interesting to explore extensions of the proposed procedure to estimate sparse normal means.

\section{\label{sec:data}Data analysis}
High-dimensional classification is an important problem in genomics. \citet{shi2010microarray} studied the effectiveness of using gene expression microarray data to develop classification rules for various phenotypes. This section focuses on classification of two of these phenotypes: estrogen receptor status and treatment response status in breast cancer patients. The training and validation datasets they used are publicly available from the Gene Expression Omnibus \citep{edgar2002gene} under accession number GSE20194.

Integrating auxiliary data may help improve classification accuracy. \citet{wang2005gene} developed a gene expression signature for distant metastasis-free survival in estrogen receptor-positive and -negative breast cancer patients. It may be possible to leverage data from \citet{wang2005gene}, publicly available under accession number GSE2034, to more accurately classify the two outcomes from \citet{shi2010microarray}. However, it is not clear how to best integrate these auxiliary data.

The normal means estimation problem using side information, studied in this paper, provides one approach. \citet{greenshtein2009application} showed that minimizing the squared error risk in the normal means problem is closely connected to minimizing the misclassification rate in high dimensional classification. Let $\bar{G}_i^Y$ denote the average expression level of gene $i$ in across all training subjects in class $Y = 0, 1$ and $\hat{s}^Y_i$ denote its estimated standard deviation. \citet{greenshtein2009application} considered classifying an observed gene expression vector $(G_1, \ldots, G_n)$ using
\begin{equation}
  \label{eq:classifier}
  I\left( \sum_{i = 1}^n \hat{\theta}_i G_i / \hat{s}_i \geq c \right)
\end{equation}
for some cutoff $c$, where $\hat{s}_i = \{ (\hat{s}^1_i)^2 / n_1 + (\hat{s}^0_i)^2 / n_0 \}^{1/2}$ and $\hat{\theta}_i$ is an estimate of the expected value of $Z_i = (\bar{G}^1_i - \bar{G}^0_i) / \hat{s}_i$. They showed that using the $f$-modeling procedure of \citet{brown2009nonparametric} to obtain $\hat{\theta}_i$  can lead to more accurate classification compared to simply using $\hat{\theta}_i = Z_i$.

Combined with ideas in this paper, this framework leads to a natural integrative classifier. Let $X_{i1}$ equal $Z_i$ calculated for either estrogen receptor status or treatment response status from the \citet{shi2010microarray} study, and let $X_{i2}$ be the differential expression $Z$-score of the $i$th gene with respect to either estrogen receptor status or distant metastasis-free survival from the \citet{wang2005gene} study. Integrating $X_{i2}$ into the estimate $\hat{\theta}_{i1}$ should lead to more accurate classification.

This integrative classification was implemented using the proposed rule $\bdelta^{\hat{\bt}}_\rho$ \eqref{eq:proposed}, the method of \citet{kou2017optimal} using a model linear in $X_{i2}$, and the procedure of \citet{banerjee2018adaptive} for sparse normal means. These were compared to five classifiers that do not make use of auxiliary information: 1) the method of \citet{greenshtein2009application} but implemented using the $g$-modeling procedure of \citet{jiang2009general}, 2) the naive Bayes classifier, 3) logistic lasso using the R package \verb|glmnet| \citep{friedman2010regularization}, 4) random forest using the R package \verb|ranger| \citep{wright2017ranger}, and 5) the regularized optimal affine discriminant analysis of \citet{fan2012road} using the R package \verb|TULIP| \citep{pan2019tulip}. Tuning parameters for lasso and the method of \citet{fan2012road} were chosen using 10-fold cross-validation while random forest was run using default parameters.

The integrative, naive Bayes, and \citet{greenshtein2009application} classifiers all assume that the $X_{id}$ are independent across $i$. For these procedures, screening was thus first performed to ensure that the magnitude of the correlation between every pair of genes in the training data was small, similar to what was done in \citet{dicker2016high}. Specifically, genes were sorted from most to least significantly associated with the outcome in the training data, with $p$-values calculated using the R package \verb|limma| \citep{smyth2005limma}. Starting from the most significant gene, any other gene with correlation greater than 0.2 in magnitude was removed from the dataset. No screening was performed for lasso, random forest, or the method of \citet{fan2012road}.

\begin{figure}
  \includegraphics[width = \textwidth]{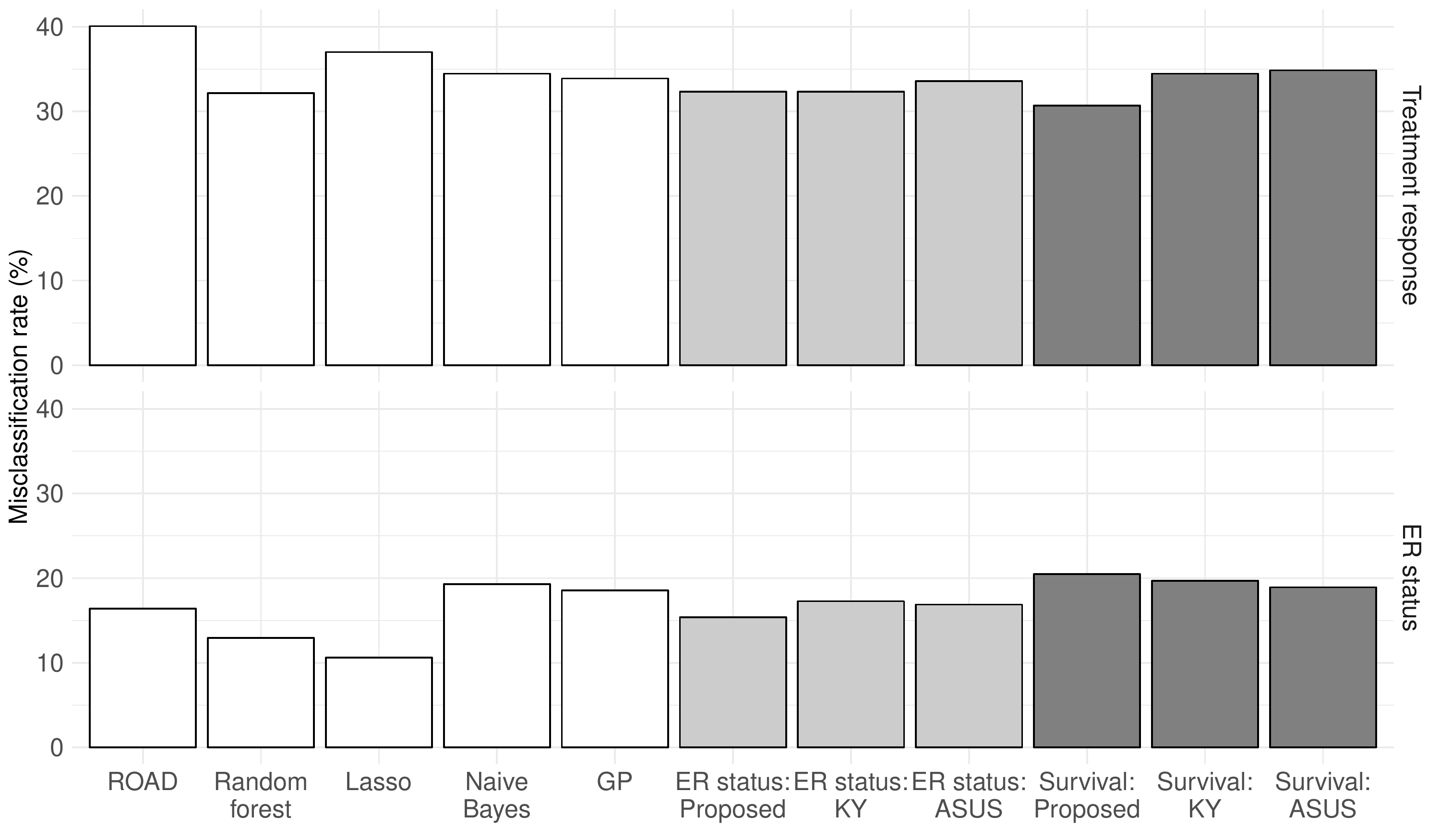}
  \caption{\label{fig:data}Average misclassification errors for treatment response status or estrogen receptor (ER) status from \citet{shi2010microarray}. GP: method of \citet{greenshtein2009application}; KY: method of \citet{kou2017optimal}; ASUS: method of \citet{banerjee2018adaptive}; ROAD: method of \citet{fan2012road}. ``+ ER status/survival'': using differential expression with respect to either ER status or distant metastasis-free survival from \citet{wang2005gene} as auxiliary data.}
\end{figure}

Misclassification rates for estrogen receptor and treatment response status were assessed using the same training and testing datasets used in \citet{shi2010microarray}, and classification was also repeated after swapping the roles of the training and testing data. The averages of the two resulting misclassification rates for the different methods are displayed in Figure \ref{fig:data}.

The results suggest that integrative classification can be a useful strategy. Intuitively, the survival results from \citet{wang2005gene} should be most informative for predicting treatment response, while the ER status data from \citet{wang2005gene} should be most useful for predicting ER status. Indeed, the proposed integrative classifier using survival $Z$-scores to predict treatment response gave the lowest misclassification rate among all methods. The proposed method integrating ER status $Z$-scores to predict ER status performed better than every method except random forest and lasso.

\section{\label{sec:disc}Discussion}
This paper assumes that the primary data $\bX_{\cdot 1}$ and the auxiliary data $\bX_{\cdot 2}$ are statistically independent. However, in some practical settings $X_{i1}$ and $X_{i2}$ may be correlated for each $i$, for example if $\bX_{\cdot 1}$ and $\bX_{\cdot 2}$ arise from case-control studies with shared controls \citep{zaykin2010p}. The ideas proposed in this paper can be naturally extended to this correlated setting. Assuming $(X_{i1}, X_{i2})$ were bivariate normal with a known correlation, the oracle integrative rule would be similar to \eqref{eq:oracle} and is given in \eqref{eq:oracle_corr} in the Appendix. An asymptotically risk-optimal data-driven estimator could then be constructed by minimizing an unbiased risk estimate derived using Stein's lemma.

As pointed out by one referee, this setting is especially interesting because when $X_{i1}$ and $X_{i2}$ are correlated, the $\bX_{\cdot 2}$ provides useful information for estimating $\btheta_{\cdot 1}$ even when $\btheta_{\cdot 2}$ and $\btheta_{\cdot 1}$ are completely unrelated. This is not true when $X_{i1}$ and $X_{i2}$ are independent. This is verified by Figure \ref{fig:sims_corr} in the Appendix, where the oracle integrative rule performed better than the oracle non-integrative rule when $\vert cor(X_{i1}, X_{i2}) \vert = 0.9$ even though $\btheta_{\cdot 2}$ was generated to be non-informative for $\btheta_{\cdot 1}$. Thus rules such as \eqref{eq:oracle} can therefore take full advantage of information about $\btheta_{\cdot 1}$ contained the auxiliary $\bX_{\cdot 2}$, whether that information comes in the form of informative $\theta_{i2}$, correlated $X_{i2}$, or both.

This paper considered only a single sequence of auxiliary data, but it is straightforward to extend the proposed procedure to multiple auxiliary sequences. However, this would result in theoretical and computational difficulties, for given $D - 1$ auxiliary datasets, Assumption \ref{a:theta} would require $\vert \theta_{id} \vert \leq n^{1/(2D) - \eta}$ for $d = 1, \ldots, D$, and the proposed procedure would require optimizing over $Dn$ parameters. It would be of great interest to study whether there exists a convex surrogate of the unbiased risk estimate \eqref{eq:sure}. An alternative approach might be to use parametric or semiparametric methods, such as those proposed by \citet{kou2017optimal}, but to endow them with data-driven model selection capabilities.

It would be interesting to extend data integration ideas to other variants of the classical normal means problem, such as heteroscedastic sequences, sparse sequences, and non-normal observed data. It would also be interesting to consider broader applications of the compound decision framework beyond the simultaneous estimation of a mean vector, such as the integrative high-dimensional classification problem in Section \ref{sec:data}.

Though this paper studied the highly stylized problem of normal means estimation with side information, its results reveal several general principles of integrative analysis. First, auxiliary data can be useful even if they are statistically independent of, and have no clearly expressible functional relationship with, primary data. The two datasets need only be related in the sense discussed in Section \ref{sec:oracle}. Second, in principle, integrating auxiliary data can only help and not harm the primary analysis. This is because it is possible to learn from the data themselves the degree to which the auxiliary data are informative, and thus the degree to which they should influence inference in the primary data. Third, nonparametric methods, such as the proposed \eqref{eq:proposed}, can asymptotically achieve ideal performance.

\section*{Acknowledgments}
The author thanks Professors Jiaying Gu, Gourab Mukherjee, and Roger Koenker for helpful discussions. This research was supported in part by NSF grant DMS-1613005.

\bibliographystyle{abbrvnat}
\bibliography{refs}

\appendix
\section{\label{sec:sims_corr}Simulations with correlated $X_{i1}$ and $X_{i2}$}
\begin{figure}[h!]
  \centering
  
  \includegraphics[width = \textwidth]{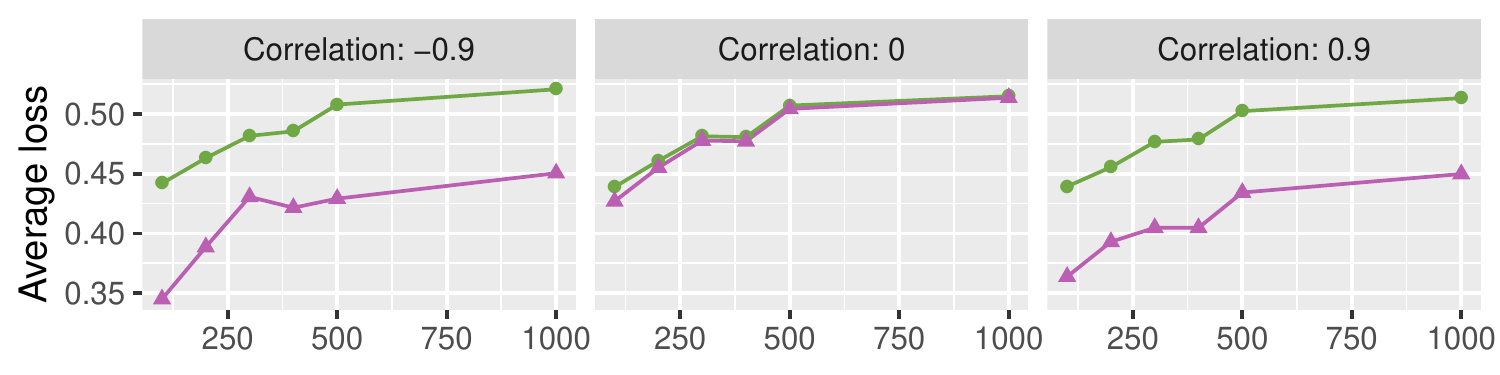}
  
  \includegraphics[width = \textwidth]{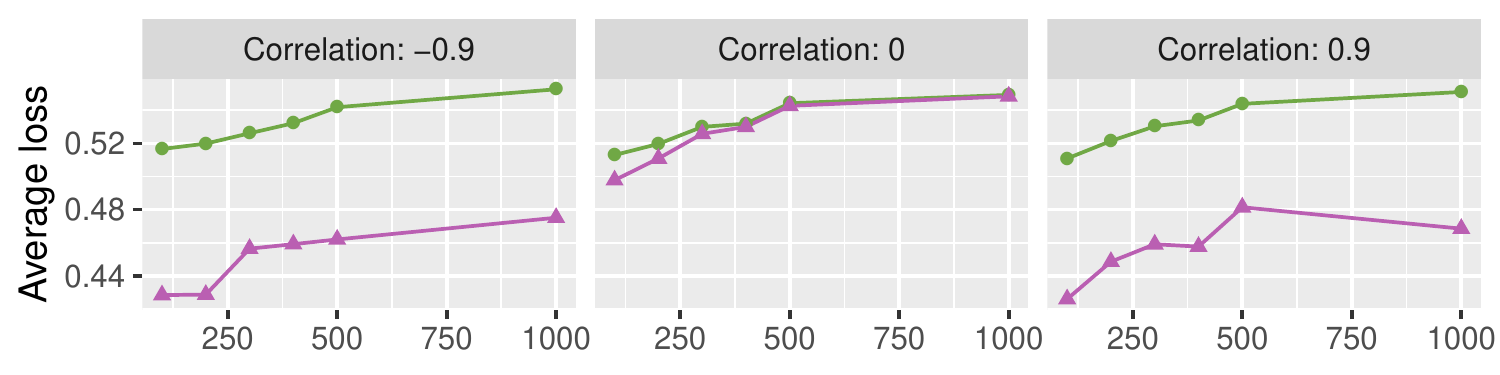}
  
  \includegraphics[width = \textwidth]{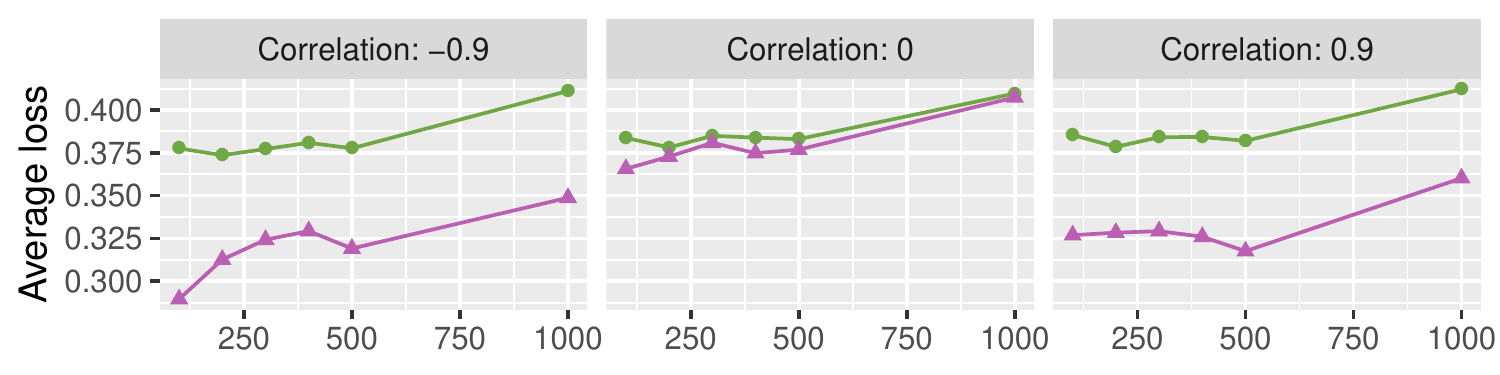}
  
  \includegraphics[width = \textwidth, trim = 0 20 0 0]{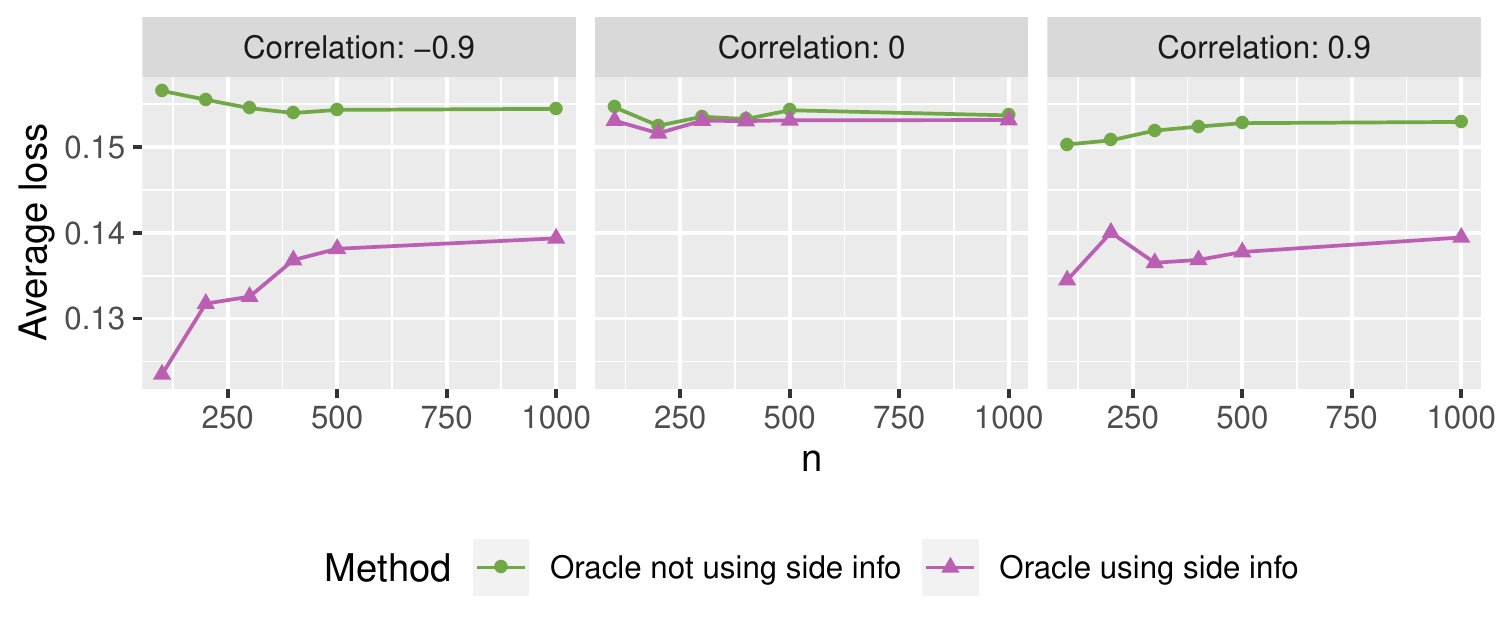} \par
  \caption{\label{fig:sims_corr}Average losses for four different configurations of $\btheta_{\cdot 1}$, non-informative $\btheta_{\cdot 2}$, and three levels of correlation between $X_{i1}$ and $X_{i2}$.}
\end{figure}

When $X_{i1}$ and $X_{i2}$ are correlated, $\bX_{\cdot 2}$ can provide useful information for estimating $\btheta_{\cdot 1}$ even if $\btheta_{\cdot 2}$ and $\btheta_{\cdot 1}$ are completely unrelated. This section illustrates this phenomenon in simulations that compare the oracle non-integrative rule
\[
\delta_i^\star(\bX_{\cdot 1})
=
\frac{\sum_{j = 1}^n \theta_{j1} \phi\{(X_{i1} - \theta_{j1}) / \sigma_1\} / \sigma_1}{\sum_{j = 1}^n \phi\{(X_{i1} - \theta_{j1}) / \sigma_1\} / \sigma_1},
\]
where $\phi(x)$ is the standard normal density, to the oracle integrative rule when $(X_{i1}, X_{i2})$ is bivariate normal with correlation $r$:
\begin{equation}
  \label{eq:oracle_corr}
  \delta_i^\star(\bX_{\cdot 1}, \bX_{\cdot 2})
  =
  \frac{\sum_{j = 1}^n \theta_{j1} \exp\{-(\bX_{i\cdot} - \btheta_{i\cdot})^\top \Sigma^{-1} (\bX_{i\cdot} - \btheta_{i\cdot}) / 2\}}{\sum_{j = 1}^n \exp\{-(\bX_{i\cdot} - \btheta_{i\cdot})^\top \Sigma^{-1} (\bX_{i\cdot} - \btheta_{i\cdot}) / 2\}},
\end{equation}
with $\bX_{i\cdot} = (X_{i1}, X_{i2})^\top$, $\btheta_{i\cdot} = (\theta_{i1}, \theta_{i2})^\top$, and
\[
\Sigma
=
\begin{pmatrix}
  \sigma_1^2 & r \sigma_1 \sigma_2 \\
  r \sigma_1 \sigma_2 & \sigma_2^2
\end{pmatrix}.
\]
The $\theta_{i1}$ and non-informative $\theta_{i2}$ were generated as in Section \ref{sec:sims}, and the $X_{id} = \theta_{id} + \epsilon_{id}$ for $d = 1, 2$, where $(\epsilon_{i1}, \epsilon_{i2})$ are bivariate normal with mean zero and variance $\Sigma$. Figure \ref{fig:sims_corr} reports the average losses of these two oracle estimators over 200 simulations for $r = -0.9, 0, 0.9$ and shows that the integrative estimator significantly outperformed the non-integrative one when $r \ne 0$.

\section{Proof of Proposition \ref{prop:oracle}}
For any separable rule $\bdelta = (\delta_1, \ldots, \delta_n) \in \mathcal{S}$ \eqref{eq:S},
\begin{align*}
  R_n(\btheta, \bdelta) =\,&
  \frac{1}{n} \sum_{i=1}^n \int \{\theta_{i1} - \delta^\star_i(x_1, x_2)\}^2
  p_i^0(x_1, x_2) dx_1 dx_2 \,+
  \\
  &\frac{2}{n} \sum_{i=1}^n \int
  \{\theta_{i1} - \delta^\star_i(x_1, x_2)\} \{\delta^\star_i(x_1, x_2) - \delta_i(x_1, x_2)\}
  p_i^0(x_1, x_2) dx_1 dx_2 \,+
  \\
  &\frac{1}{n} \sum_{i=1}^n \int
  \{\delta^\star_i(x_1, x_2) - \delta_i(x_1, x_2)\}^2
  p_i^0(x_1, x_2) dx_1 dx_2,
\end{align*}
where$ p_i^0(x_1, x_2)$ equals the density of $(X_{i1}, X_{i2})$ as defined in \eqref{eq:density}. Since $\bdelta$ and $\bdelta^\star$ are separable, the functions $\delta_i(x_1, x_2)$ and $\delta^\star_i(x_1, x_2)$ do not depend on the index $i$. Thus for the $\delta^\star_i$ given in \eqref{eq:oracle}, the middle term above equals zero, so
\begin{align*}
  R_n(\btheta, \bdelta) =\,&
  \frac{1}{n} \sum_{i=1}^n \E_{\theta} [\{\theta_{i1} - \delta^\star_i(X_{i1}, X_{i2})\}^2] +
  \frac{1}{n} \sum_{i=1}^n \E_{\theta} [\{\delta^\star_i(X_{i1}, X_{i2}) - \delta_i(X_{i1}, X_{i2})\}^2]
  \\
  \geq\,& R_n(\btheta, \bdelta^\star),
\end{align*}
for any $\bdelta \in \mathcal{S}$. Therefore $\bdelta^\star$ achieves the minimum risk within the class of separable rules.

\section{Proof of Theorem \ref{thm:regularized}}
Denoting the density of $(X_{i1}, X_{i2})$ by $p_i^0(x_1, x_2)$ as in \eqref{eq:density}, the risk of the regularized oracle rule $\bdelta^\star_\rho$ \eqref{eq:regularized} can be written as
\begin{align*}
  R_n(\btheta, \bdelta^\star_\rho) =\,&
  \frac{1}{n} \sum_{i = 1}^n \E (\theta_{i1} - X_{i1})^2 \,-
  \\
  &\frac{2}{n} \sum_{i = 1}^n \int
  \frac{
    \sum_j (\theta_{j1} - x_1) p_j^0(x_1, x_2)
  }{
    \rho + \sum_j p_j^0(x_1, x_2)
  }
  (\theta_{i1} - x_1) p_i^0(x_1, x_2) dx_1 dx_2 \,+
  \\
  &\frac{1}{n} \sum_{i = 1}^n \int
  \left\{ \frac{
    \sum_j (\theta_{j1} - x_1) p_j^0(x_1, x_2)
  }{
    \rho + \sum_j p_j^0(x_1, x_2)
  } \right\}^2
  p_i^0(x_1, x_2) dx_1 dx_2.
\end{align*}
The second term above obeys
\begin{align*}
  &
  \frac{2}{n} \sum_{i = 1}^n \int
  \frac{
    \sum_j (\theta_{j1} - x_1) p_j^0(x_1, x_2)
  }{
    \rho + \sum_j p_j^0(x_1, x_2)
  }
  (\theta_{i1} - x_1) p_i^0(x_1, x_2) dx_1 dx_2\\
  =\,&
  \frac{2}{n} \int
  \frac{
    \{ \sum_j (\theta_{j1} - x_1) p_j^0(x_1, x_2) \}^2
  }{
    \rho + \sum_j p_j^0(x_1, x_2)
  } dx_1 dx_2
  \\
  \geq\,&
  \frac{2}{n} \int
  \left\{ \frac{
    \sum_j (\theta_{j1} - x_1) p_j^0(x_1, x_2) \}^2
  }{
    \rho + \sum_j p_j^0(x_1, x_2)
  } \right\}^2 \sum_j p_j^0(x_1, x_2) dx_1 dx_2,
\end{align*}
since $\rho > 0$. Therefore
\begin{align*}
  R_n(\btheta, \bdelta^\star_\rho) \leq\,& \sigma_1^2 - \frac{1}{n} \sum_{i = 1}^n \int
  \left\{ \frac{
    \sum_j (\theta_{j1} - x_1) p_j^0(x_1, x_2)
  }{
    \rho + \sum_j p_j^0(x_1, x_2)
  } \right\}^2
  p_i^0(x_1, x_2) dx_1 dx_2.
\end{align*}
Similarly, the risk of the unregularized oracle rule $\bdelta^\star$ \eqref{eq:oracle} is
\begin{align*}
  R_n(\btheta, \bdelta^\star) =\,& \sigma_1^2 - \frac{1}{n} \sum_{i = 1}^n \int
  \left\{ \frac{
    \sum_j (\theta_{j1} - x_1) p_j^0(x_1, x_2)
  }{
    \sum_j p_j^0(x_1, x_2)
  } \right\}^2
  p_i^0(x_1, x_2) dx_1 dx_2.
\end{align*}
Since
\begin{align*}
  & \left\{ \frac{
    \sum_j (\theta_{j1} - x_1) p_j^0(x_1, x_2)
  }{
    \sum_j p_j^0(x_1, x_2)
  } \right\}^2
  -
  \left\{ \frac{
    \sum_j (\theta_{j1} - x_1) p_j^0(x_1, x_2)
  }{
    \rho + \sum_j p_j^0(x_1, x_2)
  } \right\}^2  \\
  =\,&
  \left\{ \frac{
    \sum_j (\theta_{j1} - x_1) p_j^0(x_1, x_2)
  }{
    \sum_j p_j^0(x_1, x_2)
  } \right\}^2
  \left[ 1 - \left\{ \frac{\sum_j p_j^0(x_1, x_2) }{\rho + \sum_j p_j^0(x_1, x_2) } \right\}^2 \right] \\
  =\,&
  \left\{ \frac{
    \sum_j (\theta_{j1} - x_1) p_j^0(x_1, x_2)
  }{
    \sum_j p_j^0(x_1, x_2)
  } \right\}^2
  \left\{ 1 - \frac{\sum_j p_j^0(x_1, x_2) }{\rho + \sum_j p_j^0(x_1, x_2) } \right\}
  \left\{ 1 + \frac{\sum_j p_j^0(x_1, x_2) }{\rho + \sum_j p_j^0(x_1, x_2) } \right\} \\
  \leq\,&
  \left\{ \frac{
    \sum_j (\theta_{j1} - x_1) p_j^0(x_1, x_2)
  }{
    \sum_j p_j^0(x_1, x_2)
  } \right\}^2
  \frac{2 \rho}{\rho + \sum_j p_j^0(x_1, x_2) \}},
\end{align*}
it follows that $R_n(\btheta, \bdelta^\star_\rho) - R_n(\btheta, \bdelta^\star)$ is upper bounded by
\[
\frac{2}{n} \int\left\{ \frac{
  \sum_j (\theta_{j1} - x_1) p_j^0(x_1, x_2)
}{
  \sum_j p_j^0(x_1, x_2)
} \right\}^2
\frac{\rho}{\rho + \sum_j p_j^0(x_1, x_2)}
\sum_{i = 1}^n p_i^0(x_1, x_2) dx_1 dx_2.
\]
To simplify notation, for any set $\mathcal{C} \subseteq \mathbb{R}^2$, define
\begin{equation}
  \label{eq:Delta}
  \Delta_n(\mathcal{C})
  =
  \frac{2}{n} \int_{\mathcal{C}} \left\{ \frac{
    \sum_j (\theta_{j1} - x_1) p_j^0(x_1, x_2)
  }{
    \sum_j p_j^0(x_1, x_2)
  } \right\}^2
  \frac{\rho}{\rho + \sum_j p_j^0(x_1, x_2)}
  \sum_{i = 1}^n p_i^0(x_1, x_2) dx_1 dx_2,
\end{equation}
so that $R_n(\btheta, \bdelta^\star_\rho) - R_n(\btheta, \bdelta^\star) \leq \Delta_n(\mathbb{R}^2)$. It will be shown below that $\Delta_n(\mathbb{R}^2)$ tends to zero asymptotically. Since the regularized oracle rule $\bdelta^\star_\rho$ is separable \eqref{eq:S}, $0 \leq R_n(\btheta, \bdelta^\star_\rho) - R_n(\btheta, \bdelta^\star)$ as well, by Proposition \ref{prop:oracle}. This will conclude the proof.

To upper-bound $\Delta_n(\mathbb{R}^2)$, define the set
\[
\mathcal{L}_n = \left\{ (x_1, x_2) : \log n \leq \sum_i p_i^0(x_1, x_2) \right\},
\]
so that, roughly, $\sum_i p_i^0(x_1, x_2)$ is large on $\mathcal{L}_n$. Then $\Delta_n(\mathbb{R}^2) = \Delta_n(\mathcal{L}_n) + \Delta_n(\mathcal{L}_n^c)$, and
\begin{align*}
  \Delta_n(\mathcal{L}_n) \leq\,&
  \frac{2 \rho }{\rho + \log n} \frac{1}{n} \int_{\mathcal{L}_n}
  \left\{ \frac{ \sum_j (\theta_{j1} - x_1) p_j^0(x_1, x_2)
  }{
    \sum_j p_j^0(x_1, x_2)
  } \right\}^2 \sum_j p_j^0(x_1, x_2) dx_1 dx_2\\
  \leq\,&
  \frac{2 \rho }{\rho + \log n} \frac{1}{n} \int_{\mathcal{L}_n} \sum_j (\theta_{j1} - x_1)^2 p_j^0(x_1, x_2) dx_1 dx_2
  \leq
  \frac{2 \rho}{\rho + \log n} \sigma_1^2 \rightarrow 0,
\end{align*}
where the second inequality follows from Jensen's inequality. It remains to show that $\Delta(\mathcal{L}_n^c)$ goes to zero. To this end, further define the sets
\[
\begin{aligned}
  &\mathcal{A}_1 = \{(x_1, x_2) : x_1 \leq -C n^{1/4 - \eta}\},
  \\
  &\mathcal{A}_2 = \{(x_1, x_2) : -C n^{1/4 - \eta} < x_1 \leq -C n^{1/4 - \eta}\},
  \\
  &\mathcal{A}_3 = \{(x_1, x_2) : C n^{1/4+\eta} < x_1\}.
\end{aligned}
\]
It will be shown that $\Delta_n(\mathcal{L}_n^c \cap \mathcal{A}_1)$, $\Delta_n(\mathcal{L}_n^c \cap \mathcal{A}_2)$, and $\Delta_n(\mathcal{L}_n^c \cap \mathcal{A}_3)$ converge to zero.

First,
\[
  \Delta_n(\mathcal{L}_n^c \cap \mathcal{A}_1) \leq \frac{2}{n} \int_{\mathcal{L}_n^c \cap \mathcal{A}_1}
  \frac{
    \sum_j (\theta_{j1} - x_1) p_j^0(x_1, x_2)
  }{
    \sum_j p_j^0(x_1, x_2)
  }
  \sum_{i = 1}^n (\theta_{i1} - x_1) p_i^0(x_1, x_2) dx_1 dx_2.
\]
Defining the function $g(x) = \exp\{ x / (2 \sigma_1^2) \}$, Jensen's inequality gives
\begin{align*}
  g\left\{
  \frac{
    \sum_j (\theta_{j1} - x_1)^2 p_j^0(x_1, x_2)
  }{
    \sum_j p_j^0(x_1, x_2)
  } \right\}
  \leq\,&
  \frac{
    \sum_j g\{(\theta_{j1} - x_1)^2\} p_j^0(x_1, x_2)
  }{
    \sum_j p_j^0(x_1, x_2)
  }
  \\
  =\,&
  \frac{\sum_j (2 \pi \sigma_1 \sigma_2)^{-1} \exp\{-(x_2 - \theta_{j2}) / \sigma_2^2\} }{ \sum_j p_j^0(x_1, x_2) }
  \leq
  \frac{(2 \pi \sigma_1 \sigma_2)^{-1} n}{ \sum_j p_j^0(x_1, x_2) },
\end{align*}
which is always at least 1. Combining this with another application of Jensen's inequality implies
\begin{equation}
  \label{eq:jensen}
  \left\vert
  \frac{
    \sum_j (\theta_{j1} - x_1) p_j^0(x_1, x_2)
  }{
    \sum_j p_j^0(x_1, x_2)
  }
  \right\vert
  \leq
  \left\{
  \frac{
    \sum_j (\theta_{j1} - x_1)^2 p_j^0(x_1, x_2)
  }{
    \sum_j p_j^0(x_1, x_2)
  }
  \right\}^{1/2}
  \leq
  \left\{ 2 \sigma_1^2 \log \frac{(2 \pi \sigma_1 \sigma_2)^{-1} n}{ \sum_j p_j^0(x_1, x_2) } \right\}^{1/2}.
\end{equation}
Since $\vert \theta_{j1} \vert \leq C n^{1/4 - \eta}$ by Assumption \ref{a:theta}, $\theta_{i1} - x_1 \geq 0$ on $\mathcal{A}_1$, so
\[
\Delta_n(\mathcal{L}_n^c \cap \mathcal{A}_1)
\leq\,
\frac{2}{n} \int_{\mathcal{L}_n^c \cap \mathcal{A}_1}
\left\{ 2 \sigma_1^2 \log \frac{(2 \pi \sigma_1 \sigma_2)^{-1} n}{ \sum_j p_j^0(x_1, x_2) } \right\}^{1/2}
\sum_{i = 1}^n (\theta_{i1} - x_1) p_i^0(x_1, x_2) dx_1 dx_2.
\]
Now for each value of $x_2$, define the function
\[
u_{x_2}(x_1) = n^{-1} 2 \pi \sigma_1 \sigma_2 \sum_j p_j^0(x_1, x_2)
\]
so that
\[
\frac{du_{x_2}}{dx_1} = n^{-1} 2 \pi \sigma_1 \sigma_2 \sum_j (\theta_{j1} - x_1) p_j^0(x_1, x_2) dx_1.
\]
This implies that
\begin{align*}
  &
  \Delta_n(\mathcal{L}_n^c \cap \mathcal{A}_1) \\
  \leq\,&
  \frac{1}{\pi \sigma_1 \sigma_2} \int_{-\infty}^{\infty}
  \int_{u_{x_2}(-\infty)}^{u_{x_2}(-C n^{1/4 - \eta})}
  I\{ n^{-1} 2 \pi \sigma_1 \sigma_2 \log n \geq u_{x_2}(x_1) \}
  \left( 2 \sigma_1^2 \log \frac{1}{ u_{x_2} } \right)^{1/2}
  du_{x_2} dx_2
  \\
  =\,& \frac{1}{\pi \sigma_2} \int_{-\infty}^{\infty}
\int_{u_{x_2}(-\infty)}^{u_{x_2}(-C n^{1/4 - \eta})}
  I\{ n^{-1} 2 \pi \sigma_1 \sigma_2 \log n \geq u_{x_2}(x_1) \}
  \left( \log \frac{1}{ u_{x_2} } \right)^{1/2}
  du_{x_2} dx_2
\end{align*}
Since $\theta_{j1} - x_1 > 0$ for all $j$ on $\mathcal{A}_1$. Therefore, $u_{x_2}(x_1)$ is increasing on $\mathcal{A}_1$ and $u_{x_2}(-C n^{1/4 - \eta}) > 0$. Then
\begin{align*}
  \Delta_n(\mathcal{L}_n^c \cap \mathcal{A}_1) \leq\,&
  \frac{1}{\pi \sigma_2} \int_{-\infty}^{\infty} \int_0^{n^{-1} 2 \pi \sigma_1 \sigma_2 \log n\wedge u_{x_2}(-C n^{1/4 - \eta})}
  \left( \log \frac{1}{ u_{x_2}^2 } \right)^{1/2}
  du_{x_2} dx_2
  \\
  \leq\,& \frac{1}{\pi \sigma_2} \int_{-\infty}^{\infty} 
  \left\{ \int_0^{u_{x_2}(-C n^{1/4 - \eta})} 1 du_{x_2} \right\}^{1/2}
  \left\{ \int_0^{n^{-1} 2 \pi \sigma_1 \sigma_2 \log n} \log \frac{1}{u_{x_2}^2} du_{x_2} \right\}^{1/2}
  dx_2,
\end{align*}
where the last line follows from the Cauchy-Schwarz inequality. Integrating by parts gives
\begin{align*}
  \int_0^{n^{-1} 2 \pi \sigma_1 \sigma_2 \log n} \log \frac{1}{u_{x_2}^2} du_{x_2} =\,&
  u_{x_2} \log \frac{1}{u_{x_2}^2} \bigg\vert_0^{n^{-1} 2 \pi \sigma_1 \sigma_2 \log n} +
  \int_0^{n^{-1} 2 \pi \sigma_1 \sigma_2 \log n} 2 du_{x_2} dx_2
  \\
  =\,& \frac{2 \pi \sigma_1 \sigma_2 \log n}{n} \left( 2 \log \frac{n}{2 \pi \sigma_1 \sigma_2 \log n} + 2 \right),
\end{align*}
so
\[
\Delta_n(\mathcal{L}_n^c \cap \mathcal{A}_1) \leq
\frac{2 \sigma_1 \log^{1/2} n}{n} \left( 2 \log \frac{n}{2 \pi \sigma_1 \sigma_2 \log n} + 2 \right)^{1/2}
\int_{-\infty}^\infty \left\{ \sum_j p_j^0(-Cn^{1/4 - \eta}, x_2) \right\}^{1/2} dx_2.
\]

Finally, first integrate the remaining integral over $x_2 \leq -Cn^{1/4 - \eta}$. Since $\vert \theta_{j2} \vert \leq C n^{1/4 - \eta}$ by Assumption \ref{a:theta}, it follows that $\theta_{j2} - x_2 \geq -C n^{1/4 - \eta} - x_2 \geq 0$ for all $j = 1, \ldots, n$ and thus that $(\theta_{j2} - x_2)^2 \geq (-C n^{1/4 - \eta} - x_2)^2$. Therefore
\begin{align*}
  \sum_j p_j^0(-Cn^{1/4 - \eta}, x_2) =\,&
  \frac{1}{2 \pi \sigma_1 \sigma_2} \sum_j \exp\left\{-\frac{1}{2 \sigma_1^2} (-Cn^{1/4 - \eta} - \theta_{j1})^2 \right\} \exp\left\{-\frac{1}{2 \sigma_2^2} (x_2 - \theta_{j2})^2 \right\}
  \\
  \leq\,& \frac{n}{2 \pi \sigma_1 \sigma_2} \exp\left\{-\frac{1}{2 \sigma_2^2} (x_2 + C n^{1/4 - \eta})^2 \right\},
\end{align*}
which implies that
\begin{align*}
  & \int_{-\infty}^{-C n^{1/4 - \eta}} \left\{ \sum_j p_j^0(-Cn^{1/4 - \eta}, x_2) \right\}^{1/2} dx_2
  \\
  \leq \,& \int_{-\infty}^{-C n^{1/4 - \eta}} \frac{n^{1/2}}{(2 \pi \sigma_1 \sigma_2)^{1/2}} \exp\left\{-\frac{1}{4 \sigma_2^2} (x_2 + C n^{1/4 - \eta})^2 \right\} dx_2
  \\
  \leq\,& \frac{(2 \sigma_2 n)^{1/2}}{\sigma_1^{1/2}} \int_{-\infty}^{-C n^{1/4 - \eta}} \frac{1}{(2 \pi 2 \sigma_2^2)^{1/2}} \exp\left\{-\frac{1}{4 \sigma_2^2} (x_2 + C n^{1/4 - \eta})^2 \right\} dx_2
  \leq \frac{(2 \sigma_2 n)^{1/2}}{\sigma_1^{1/2}}.
\end{align*}
A similar calculation shows that on $Cn^{1/4 - \eta} < x_2$,
\begin{align*}
  \int_{C n^{1/4 - \eta}}^\infty \left\{ \sum_j p_j^0(-Cn^{1/4 - \eta}, x_2) \right\}^{1/2} dx_2 \leq \frac{(2 \sigma_2 n)^{1/2}}{\sigma_1^{1/2}}
\end{align*}
as well. Finally, on $-Cn^{1/4 - \eta} < x_2 \leq Cn^{1/4 - \eta}$,
\begin{align*}
  \int_{-C n^{1/4 - \eta}}^{C n^{1/4 - \eta}} \left\{ \sum_j p_j^0(-Cn^{1/4 - \eta}, x_2) \right\}^{1/2} dx_2 \leq \frac{n^{1/2}}{(2 \pi \sigma_1 \sigma_2)^{1/2}} 2C n^{1/4 - \eta}.
\end{align*}
Putting these results together shows that $\Delta_n(\mathcal{L}_n^c \cap \mathcal{A}_1)$ is at most
\begin{align*}
  \frac{2 \sigma_1 \log^{1/2} n}{n} \left( 2 \log \frac{n}{2 \pi \sigma_1 \sigma_2 \log n} + 2 \right)^{1/2}
  \left\{ 2\frac{(2 \sigma_2 n)^{1/2}}{\sigma_1^{1/2}} + \frac{n^{1/2}}{(2 \pi \sigma_1 \sigma_2)^{1/2}} 2C n^{1/4 - \eta} \right\}
  \rightarrow 0.
\end{align*}
A very similar calculation can be used to show that $\Delta_n(\mathcal{L}_n^c \cap \mathcal{A}_3) \rightarrow 0$ as well.

It remains to bound $\Delta_n(\mathcal{L}_n^c \cap \mathcal{A}_2)$. Denote $u(x_1, x_2) = \sum_j p_j^0(x_1, x_2)$ and use \eqref{eq:jensen} see that $\Delta_n(\mathcal{L}_n^c \cap \mathcal{A}_2)$ is at most
\begin{align*}
  \frac{4 \sigma_1^2}{n} \int_{-\infty}^\infty \int_{-C n^{1/4 - \eta}}^{C n^{1/4 - \eta}} I\{ \log n \geq u(x_1, x_2) \}
  \log \frac{(2 \pi \sigma_1 \sigma_2)^{-1} n}{u(x_1, x_2)} u(x_1, x_2) dx_1 dx_2.
\end{align*}
Now for any constant $K$,
\[
\frac{d}{du} \left( u^{1/2} \log \frac{K}{u} \right) = \frac{1}{u^{1/2}} \left( \frac{1}{2} \log \frac{K}{u} - 1 \right).
\]
This implies that the function $u^{1/2} \log (K / u)$ increases for $u \leq K e^{-2}$ and decreases thereafter. Now consider the function $2 u^{1/2}$, which equals $u^{1/2} \log (K / u)$ at $u = K e^{-2}$ and increases thereafter. Therefore $u^{1/2} \log (K / u)$ is upper-bounded by the nondecreasing function $u^{1/2} \{ 2 \vee \log (K / u) \}$. Therefore $\Delta_n(\mathcal{L}_n^c \cap \mathcal{A}_2)$ is at most
\begin{align*}
  &\frac{4 \sigma_1^2}{n} \int_{-\infty}^\infty \int_{-C n^{1/4 - \eta}}^{C n^{1/4 - \eta}} I\{ \log n \geq u(x_1, x_2) \}
  \left\{ 2 \vee \log \frac{(2 \pi \sigma_1 \sigma_2)^{-1} n}{u(x_1, x_2)} \right\} u(x_1, x_2) dx_1 dx_2 \\
  \leq\,& \frac{4 \sigma_1^2 \log^{1/2} n}{n} \left\{ 2 \vee \log \frac{(2 \pi \sigma_1 \sigma_2)^{-1} n}{\log n} \right\}
  \int_{-\infty}^\infty \int_{-C n^{1/4 - \eta}}^{C n^{1/4 - \eta}} u^{1/2}(x_1, x_2) dx_1 dx_2\\
  \leq\,& \frac{4 \sigma_1^2 \log^{1/2} n}{n} \left\{ 2 \vee \log \frac{(2 \pi \sigma_1 \sigma_2)^{-1} n}{\log n} \right\} \frac{2 C n^{1/4 - \eta}}{(2 \pi \sigma_1 \sigma_2)^{1/2}} \times \\
  & \int_{-\infty}^\infty \left[ \sum_j \exp\left\{-\frac{1}{2 \sigma_2^2} (x_2 - \theta_{j2})^2 \right\} \right]^{1/2} dx_2.
\end{align*}

To finish bounding $\Delta_n(\mathcal{L}_n^c \cap \mathcal{A}_2)$, integrate over $x_2 \leq -C n^{1/4 - \eta}$, $-C n^{1/4 - \eta} < x_2 \leq C n^{1/4 - \eta}$, and $C n^{1/4 - \eta} < x_2$, as was done when bounding $\Delta_n(\mathcal{L}_n^c \cap \mathcal{A}_1)$. Using Assumption \ref{a:theta}, it can be shown that
\begin{align*}
  &\int_{-\infty}^\infty \left[\sum_j \exp\left\{-\frac{1}{2 \sigma_2^2} (x_2 - \theta_{j2})^2 \right\} \right]^{1/2} dx_2
  \leq n^{1/2} (4 \pi^{1/2} \sigma_2 + 2 C n^{1/4 - \eta}),
\end{align*}
which means that
\[
\Delta_n(\mathcal{L}_n^c \cap \mathcal{A}_2)
\leq
\frac{4 \sigma_1^2 \log^{1/2} n}{n^{1/2}} \left\{ 2 \vee \log \frac{(2 \pi \sigma_1 \sigma_2)^{-1} n}{\log n} \right\} \frac{2 C n^{1/4 - \eta}}{(2 \pi \sigma_1 \sigma_2)^{1/2}} (4 \pi^{1/2} \sigma_2 + 2 C n^{1/4 - \eta})
\rightarrow 0.
\]

\section{Proof of Theorem \ref{thm:uniform}}
\subsection{Outline}
For the parameter vector $\bt = (t_{11}, \ldots, t_{n1}, t_{12}, \ldots, t_{n2})$, define the function
\begin{equation}
  \label{eq:f}
  \begin{aligned}
    f(x_1, x_2, \mu; \bt)
    =\, &
    \frac{
      \sum_j (t_{j1} - x_1)^2 p(x_1, x_2; t_{j1}, t_{j2})
    }{
      \rho + \sum_j p(x_1, x_2; t_1, t_2)
    }
    -
    \sigma_1^2 \frac{
      \sum_j p(x_1, x_2; t_{j1}, t_{j2})
    }{
      \rho + \sum_j p(x_1, x_2; t_{j1}, t_{j2})
    }
    \,- \\
    &
    \left\{
    \frac{
      \sum_j (t_{1j} - x_1) p(x_1, x_2; t_{j1}, t_{j2})
    }{
      \rho + \sum_j p(x_1, x_2; t_{j1}, t_{j2})
    }
    \right\}^2
    -
    (x_1 - \mu)
    \frac{
      \sum_j (t_{1j} - x_1) p(x_1, x_2; t_{j1}, t_{j2})
    }{
      \rho + \sum_j p(x_1, x_2; t_{j1}, t_{j2})
    }.
  \end{aligned}
\end{equation}
Then it is straightforward to show that
\begin{align*}
  \textsc{sure}(\bt) - \ell_n(\bt) = \sigma_1^2 - \frac{1}{n} \sum_{i = 1}^n (X_{i1} - \theta_{i1})^2 +
  \frac{2}{n} \sum_{i = 1}^n f(X_{i1}, X_{i2}, \theta_{i1}; \bt).
\end{align*}

By Jensen's inequality,
\begin{align*}
  \left \{ E \left \vert \sigma_1^2 - \frac{1}{n} \sum_i (X_{i1} - \theta_{i1})^2 \right \vert \right \}^2
  \leq\,&
  E \left \{ \sigma_1^2 - \frac{1}{n} \sum_i (X_{i1} - \theta_{i1})^2 \right \}^2 \\
  =\,&
  \sigma_1^4 - \frac{2\sigma_1^2}{n} \sum_i E(X_{i1} - \theta_{i1})^2 + \frac{1}{n^2} E \left \{ \sum_i (X_{i1} - \theta_{i1})^2 \right \}^2 \\
  =\,&
  -\sigma_1^4 + \frac{\sigma_1^4}{n^2} E \left \{ \sum_i \left( \frac{X_{i1} - \theta_{i1}}{\sigma_1} \right)^2 \right \}^2.
\end{align*}
Next, the variable $\sum_i (X_{i1} - \theta_{i1})^2 / \sigma_1^2$ has a $\chi^2_n$ distribution, which has mean $n$ and variance $2n$. Therefore
\[
\left \{ E \left \vert \sigma_1^2 - \frac{1}{n} \sum_i (X_{i1} - \theta_{i1})^2 \right \vert \right \}^2
\leq
-\sigma_1^4 + \frac{\sigma_1^4}{n^2} (n^2 + 2n) = \frac{2 \sigma_1^4}{n} \rightarrow 0.
\]
The remainder of the proof uses empirical process techniques to show that
\[
E \sup_{\bt \in \mathcal{T}} \left \vert \frac{1}{n} \sum_i f(X_{i1}, X_{i2}, \theta_{i1}; \bt) \right \vert \rightarrow 0.
\]

\subsection{Symmetrization}
The terms $f(X_{i1}, X_{i2}, \theta_{i1}; \bt)$ are independent but not identically distributed. Nevertheless, the proof of the symmetrization Lemma 2.3.1 of \citet{vandervaart1996weak} still holds and implies that
\[
E \sup_{\bt \in \mathcal{T}} \left \vert \frac{1}{n} \sum_i \left\{ f(X_{i1}, X_{i2}, \theta_{i1}; \bt) - E f(X_{i1}, X_{i2}, \theta_{i1}; \bt) \right\} \right \vert
\leq
2 E \sup_{\bt \in \mathcal{T}} \left \vert \frac{1}{n} \sum_i \varepsilon_i f(X_{i1}, X_{i2}, \theta_{i1}; \bt) \right \vert,
\]
where the $\varepsilon_i$ are Rademacher random variables independent of the $X_{i1}$ and $X_{i2}$. Using \citet{stein1981estimation}, it can be shown that $E f(X_{i1}, X_{i2}, \theta_{i1}; \bt) = 0$. Therefore it only remains to show that
\begin{equation}
  \label{eq:symm}
  E \sup_{\bt \in \mathcal{T}} \left \vert \frac{1}{n} \sum_i \varepsilon_i f(X_{i1}, X_{i2}, \theta_{i1}; \bt) \right \vert \rightarrow 0.
\end{equation}

\subsection{Truncation}
This section first constructs a function $F_n(x_1, \mu)$ that satisfies $\vert f(x_1, x_2, \mu; \bt) \vert \leq F_n(x_1, \mu)$ as well as $E\{ F_n(X_{i1}, \mu) \} < \infty$, for $f$ defined in \eqref{eq:f}. By the triangle inequality,
\begin{align*}
  & \vert f(x_1, x_2, \mu; \bt) \vert \\
  \leq\, &
  \frac{
    \sum_j (t_{j1} - x_1)^2 p(x_1, x_2; t_{j1}, t_{j2})
  }{
    \rho + \sum_j p(x_1, x_2; t_1, t_2)
  }
  +
  \sigma_1^2
  \,+ \\
  &
  \left\{ \frac{
    \sum_j (t_{1j} - x_1) p(x_1, x_2; t_{j1}, t_{j2})
  }{
    \rho + \sum_j p(x_1, x_2; t_{j1}, t_{j2})
  } \right\}^2
  +
  \vert x_1 - \mu \vert
  \left \vert \frac{
    \sum_j (t_{1j} - x_1) p(x_1, x_2; t_{j1}, t_{j2})
  }{
    \rho + \sum_j p(x_1, x_2; t_{j1}, t_{j2})
  } \right \vert.
\end{align*}
Using Jensen's inequality as in \eqref{eq:jensen}, it follows that
\begin{align*}
  \frac{
    \sum_j (t_{j1} - x_1)^2 p(x_1, x_2; t_{j1}, t_{j2})
  }{
    \rho + \sum_j p(x_1, x_2; t_1, t_2)
  }
  \leq\,&
  \frac{\sum_j p(x_1, x_2; t_1, t_2) }{ \rho + \sum_j p(x_1, x_2; t_1, t_2) }
  2 \sigma_1^2 \log \frac{ (2 \pi \sigma_1 \sigma_2)^{-1} n }{ \sum_j p(x_1, x_2; t_1, t_2) }.
\end{align*}
Let $y = p(x_1, x_2; t_1, t_2)$ and $C = (2 \pi \sigma_1 \sigma_2)^{-1} n$ such that this upper bound can be written as
\[
u(y) = \frac{y}{\rho + y} 2 \sigma_1^2 \log \frac{C}{y},
\]
where $0 \leq y \leq C$ by the definition of $p(x_1, x_2; t_1, t_2)$ in \eqref{eq:density}. Since
\[
u'(y) = \frac{2 \sigma_1^2}{\rho + y} \left( \frac{\rho}{\rho + y} \log \frac{C}{y} - 1 \right)
\]
and the function $\{\rho / (\rho + y) \} \log (C / y) - 1$ is monotone decreasing, $u(y)$ attains a unique maximum at a $y^\star$ that satisfies $u'(y^\star) = 0$. This implies that
\[
u(y)
\leq
\frac{2 \sigma_1^2 y^\star}{\rho} \frac{\rho}{\rho + y^\star} \log \frac{C}{y^\star}
=
\frac{2 \sigma_1^2 y^\star}{\rho}
\]
for all $0 \leq y \leq C$. The exact value of $y^\star$ is difficult to determine exactly, by assumption $0 < \rho \leq 1$, so for every $n \geq 2 \pi \sigma_1 \sigma_2 e^{1/e}$, $\rho + \log C \geq 0$, $\log \log C \geq -1$, and
\[
u'(\log C)
=
\frac{2 \sigma_2^2}{\rho + \log C} \frac{\rho \log C - \rho \log \log C - \rho - \log C}{\rho + \log C}
\leq
\frac{2 \sigma_2^2}{\rho + \log C} \frac{(\rho - 1) \log C}{\rho + \log C}
\leq
0.
\]
Since $u'(y^\star) = 0$, it follows that $y^\star \leq \log C$ and $u(y) \leq (2 \sigma_1^2 \log C) / \rho$, or in other words,
\begin{equation}
  \label{eq:jensen_rho}
  \frac{
    \sum_j (t_{j1} - x_1)^2 p(x_1, x_2; t_{j1}, t_{j2})
  }{
    \rho + \sum_j p(x_1, x_2; t_1, t_2)
  }
  \leq
  \frac{2 \sigma_1^2}{\rho} \log \frac{n}{2 \pi \sigma_1 \sigma_2}.
\end{equation}
Furthermore, by Jensen's inequality
\begin{align*}
  \left\{ \frac{
    \sum_j (t_{j1} - x_1) p(x_1, x_2; t_{j1}, t_{j2})
  }{
    \rho + \sum_j p(x_1, x_2; t_1, t_2)
  } \right\}^2
  \leq
  \frac{
    \sum_j (t_{j1} - x_1)^2 p(x_1, x_2; t_{j1}, t_{j2})
  }{
    \rho + \sum_j p(x_1, x_2; t_1, t_2)
  }.
\end{align*}
Therefore $\vert f(x_1, x_2, \mu; \bt) \vert \leq F_n(x_1, \mu)$ for the function
\begin{equation}
  \label{eq:F_n}
  F_n(x_1, \mu) = \sigma_1^2 + \frac{4 \sigma_1^2}{\rho} \log \frac{n}{2 \pi \sigma_1 \sigma_2} + \frac{2^{1/2} \sigma_1}{\rho^{1/2}} \vert x_1 - \mu \vert \log^{1/2} \frac{n}{2 \pi \sigma_1 \sigma_2}.
\end{equation}

The next step to truncate $f$ \eqref{eq:f} in expression \eqref{eq:symm}, which will be useful when applying a maximal inequality later in the proof. Define
\begin{equation}
  \label{eq:C_n}
  C_n = (2 \sigma_1^2 \log \log n)^{1/2}.
\end{equation}
By the triangle inequality and the construction of the function $F_n$ \eqref{eq:F_n},
\begin{align*}
  E \sup_{\bt \in \mathcal{T}} \left \vert \frac{1}{n} \sum_i \varepsilon_i f(X_{i1}, X_{i2}, \theta_{i1}; \bt) \right \vert
  \leq\,&
  E \sup_{\bt \in \mathcal{T}} \left \vert \frac{1}{n} \sum_i \varepsilon_i f(X_{i1}, X_{i2}, \theta_{i1}; \bt) I(\vert X_{i1} - \theta_{i1}\vert \leq C_n )\right \vert \,+ \\
  &\frac{1}{n} \sum_i E\{ F_n(X_{i1}, \theta_{i1}) I(\vert X_{i1} - \theta_{i1}\vert > C_n) \}.
\end{align*}
For each $i$, $E\{ F_n(X_{i1}, \theta_{i1}) I(\vert X_{i1} - \theta_{i1}\vert > C_n) \}$ equals
\begin{align*}
  &\left( \sigma_1^2 + \frac{4 \sigma_1^2}{\rho} \log \frac{n}{2 \pi \sigma_1 \sigma_2} \right) P(\vert X_{i1} - \theta_{i1}\vert > C_n) +\, \\
  &\frac{2^{1/2} \sigma_1}{\rho^{1/2}} E \{ \vert X_{i1} - \theta_{i1} \vert I(\vert X_{i1} - \theta_{i1}\vert > C_n) \} \log^{1/2} \frac{n}{2 \pi \sigma_1 \sigma_2} 
\end{align*}
By Mill's inequality,
\begin{align*}
  P(\vert X_{i1} - \theta_{i1}\vert > C_n)
  =\,&
  P(\sigma_1^{-1} \vert X_{i1} - \theta_{i1}\vert  > \sigma_1^{-1} C_n)
  \leq
  \frac{2^{1/2} \sigma_1}{\pi^{1/2} C_n \log n}
\end{align*}
and furthermore,
\begin{align*}
  E \{ \vert X_{i1} - \theta_{i1} \vert I(\vert X_{i1} - \theta_{i1}\vert > C_n) \}
  =\,&
  \int \vert z \vert I(\vert z \vert > C_n) \frac{1}{(2 \pi \sigma_1^2)^{1/2}} \exp\left(-\frac{1}{2\sigma_1^2} z^2 \right) dz\\
  =\,&
  -\frac{2 \sigma_1}{(2 \pi)^{1/2}} \int_{C_n}^\infty \left(-\frac{z}{\sigma_1^2}\right) \exp\left(-\frac{1}{2\sigma_1^2} z^2 \right) dz \\
  =\,&
  \frac{2 \sigma_1}{(2 \pi)^{1/2} \log n}.
\end{align*}
This implies that $n^{-1} \sum_i E\{ F_n(X_{i1}, \theta_{i1}) I(\vert X_{i1} - \theta_{i1}\vert > C_n) = O(C_n^{-1}) \rightarrow 0$, and it remains to show that
\begin{equation}
  \label{eq:trunc}
  E \sup_{\bt \in \mathcal{T}} \left \vert \frac{1}{n} \sum_i \varepsilon_i f(X_{i1}, X_{i2}, \theta_{i1}; \bt) I(\vert X_{i1} - \theta_{i1}\vert \leq C_n ) \right \vert \rightarrow 0.
\end{equation}

\subsection{Approximation and covering number}
For $C n^{1/4 - \eta}$ as in Assumption \ref{a:theta} and some integer $K_n$ that can grow with $n$, define
\begin{equation}
  \label{eq:sets}
  \begin{aligned}
    \mathcal{U}(K_n) =\,& \left\{(u_{11}, \ldots, u_{K_n1}, u_{12}, \ldots, u_{K_n2}) : \vert u_{kd} \vert \leq C n^{1/4 - \eta}, k = 1, \ldots, K_n, d = 1, 2 \right\},\\
    \mathcal{W}(K_n) =\,& \left\{(w_1, \ldots, w_{K_n}) : w_k > 0, \sum_k w_k = 1  \right\}.
  \end{aligned}
\end{equation}
Next define functions
\begin{equation}
  \label{eq:g}
  \begin{aligned}
    &g(x_1, x_2, \mu; \bu, \bw)\\
    =\, &
    \frac{
      \sum_{k = 1}^{K_n} (u_{k1} - x_1)^2 p(x_1, x_2; u_{k1}, u_{k2}) w_k
    }{
      \rho n^{-1} + \sum_{k = 1}^{K_n} p(x_1, x_2; u_{k1}, u_{k2}) w_k
    }
    -
    \sigma_1^2 \frac{
      \sum_{k = 1}^{K_n} p(x_1, x_2; u_{k1}, u_{k2})
    }{
      \rho n^{-1} + \sum_{k = 1}^{K_n} p(x_1, x_2; u_{k1}, u_{k2}) w_k
    }
    \,- \\
    &
    \left\{
    \frac{
      \sum_{k = 1}^{K_n} (u_{k1} - x_1) p(x_1, x_2; u_{k1}, u_{k2}) w_k
    }{
      \rho n^{-1} + \sum_{k = 1}^{K_n} p(x_1, x_2; u_{k1}, u_{k2}) w_k
    }
    \right\}^2
    -
    \vert x - \mu \vert
    \frac{
      \sum_{k = 1}^{K_n} (u_{k1} - x_1) p(x_1, x_2; u_{k1}, u_{k2}) w_k
    }{
      \rho n^{-1} + \sum_{k = 1}^{K_n} p(x_1, x_2; u_{k1}, u_{k2}) w_k
    }
  \end{aligned}
\end{equation}
indexed by $\bu \in \mathcal{U}(K_n)$ and $\bw \in \mathcal{W}(K_n)$. This section constructs discrete subsets $\mathbb{U}(K_n) \subset \mathcal{U}(K_n)$ and $\mathbb{W}(K_n) \subset \mathcal{W}(K_n)$ such that for all $\bt \in \mathcal{T}$ \eqref{eq:T}, there exists some $(\bu, \bw) \in \mathbb{U}(K_n) \times \mathbb{W}(K_n)$ such that conditional on the $(X_{i1}, X_{i2})$, for any $\epsilon > 0$,
\begin{equation}
  \label{eq:approxf}
  \frac{1}{n} \sum_{i = 1}^n \vert f(X_{i1}, X_{i2}, \theta_{i1}; \bt) - g(X_{i1}, X_{i2}, \theta_{i1}; \bu, \bw) \vert \leq \epsilon
\end{equation}
for $n$ sufficiently large, with $C_n$ defined in \eqref{eq:C_n}.

The $\epsilon$-covering number $\mathcal{N}(\mathcal{T})$ of the set of functions $f(x_1, x_2, \mu; \bt)$ indexed by $\bt \in \mathcal{T}$ is defined to be the cardinality of the smallest set $\mathbb{U}(K_n) \times \mathbb{W}(K_n)$ such that \eqref{eq:approxf} holds. This section establishes an upper bound for this covering number, which will be a function of the $(X_{i1}, X_{i2})$ and $K_n$. The proof makes use of the following lemmas.
\begin{lemma}
  \label{lem:n_to_K}
  For any $\gamma_n \leq e^{-e}$ and any integer $q > 0$, define
  \begin{equation}
    \label{eq:MLK}
    \begin{aligned}
      M_n =\,& \max_{j,d} \frac{\vert X_{jd} \vert + C n^{1/4 - \eta}}{\sigma_d},\\
      L_n =\,& \left(1 + \frac{M_n^2 e}{2}\right) \log \frac{1}{\gamma_n},\\
      K_n =\,& (2 L_n - 2 + q) (2 L_n - 2) + 1.
    \end{aligned}
  \end{equation}
  Then for each $\bt \in \mathcal{T}$ \eqref{eq:T} there exist $K_n$ support points $\bmu \in \mathcal{U}(K_n)$ and $\bomega \in \mathcal{W}(K_n)$, with $\mathcal{U}(K_n)$ and $\mathcal{W}(K_n)$ defined in \eqref{eq:sets}, such that for each $(X_{i1}, X_{i2})$,
  \begin{align*}
    &\left \vert \frac{1}{n} \sum_{j = 1}^n (t_{j1} - X_{i1})^q p(X_{i1}, X_{i2}; t_{j1}, t_{j2}) - \sum_{k = 1}^{K_n} (\mu_{k1} - X_{i1})^q p(X_{i1}, X_{i2}; \mu_{k1}, \mu_{k2}) \omega_k \right \vert \\
    \leq\,&
    \frac{\gamma_n + 2}{\pi \sigma_1 \sigma_2} \sigma_1^q M_n^q \gamma_n.
  \end{align*}
\end{lemma}
\textit{Proof.} By Taylor expansion and because $L_n! \geq L_n^{L_n} e^{-L_n}$,
\[
\left \vert \exp\left\{-\frac{1}{2 \sigma_d^2} (X_{id} - t_{jd})^2 \right\} - \sum_{l = 0}^{L_n - 1} \frac{(-1)^l (X_{id} - t_{jd})^{2l}}{(2 \sigma_d^2)^l l!} \right \vert
\leq
\frac{\vert X_{id} - t_{jd} \vert^{2 L_n}}{(2 \sigma_d^2)^{L_n} L_n!}
\leq
\left( \frac{\vert X_{id} - t_{jd} \vert^2 e}{2 \sigma_d^2 L_n} \right)^{L_n}
\]
for $d = 1, 2$. By assumption, $\bt \in \mathcal{T}$, so
\[
\frac{\vert X_{id} - t_{jd} \vert^2 e}{2 \sigma_d^2 L_n}
\leq
\frac{M_n^2 e}{2 L_n}.
\]
Then choosing $L_n$ as in \eqref{eq:MLK} guarantees that
\begin{align*}
  \log \left( \frac{M_n^2 e}{2 \sigma_d^2 L_n} \right)^{L_n}
  \leq\,&
  \left(1 + \frac{M_n}{2}\right) \log \frac{1}{\gamma_n} \left(- \log \log \frac{1}{\gamma_n} \right)
  \leq
  \log \gamma_n,
\end{align*}
where the last inequality follows because since $\gamma_n < e^{-e}$ by assumption, $\log \gamma_n < 0$ while $\log \log \gamma_n^{-1} \geq 1$. Therefore
\[
\left \vert \exp\left\{-\frac{1}{2 \sigma_d^2} (X_{id} - t_{jd})^2 \right\} - \sum_{l = 0}^{L_n - 1} \frac{(-1)^l (X_{id} - t_{jd})^{2l}}{(2 \sigma_d^2)^l l!} \right \vert
\leq
\gamma_n,
\]
which also implies that
\[
\left \vert \sum_{l = 0}^{L_n - 1} \frac{(-1)^l (X_{id} - t_{jd})^{2l}}{(2 \sigma_d^2)^l l!} \right \vert
\leq
\gamma_n + 1.
\]
This imply that
\begin{align}
  &
  \left \vert \frac{1}{n} \sum_{j = 1}^n (t_{j1} - X_{i1})^q p(X_{i1}, X_{i2}; t_{j1}, t_{j2}) -
  \frac{1}{n} \sum_{j = 1}^n \frac{(t_{j1} - X_{i1})^q}{2 \pi \sigma_1 \sigma_2} \prod_{d = 1}^2 \sum_{l = 0}^{L_n - 1} \frac{(-1)^l (X_{id} - t_{jd})^{2l}}{(2 \sigma_d^2)^l l!}
  \right \vert \nonumber\\
  \leq\,&
  \frac{1}{n} \sum_{j = 1}^n \frac{\vert t_{j1} - X_{i1} \vert^q}{2 \pi \sigma_1 \sigma_2}
  \left \vert \exp\left\{-\frac{1}{2 \sigma_1^2} (X_{i1} - t_{j1})^2 \right\}
  -
  \sum_{l = 0}^{L_n - 1} \frac{(-1)^{l} (X_{i1} - t_{j1})^{2l}}{(2 \sigma_1^2)^l l!}
  \right \vert \,+ \nonumber\\
  & \frac{1}{n} \sum_{j = 1}^n  \frac{\vert t_{j1} - X_{i1} \vert^q (\gamma_n + 1)}{2 \pi \sigma_1 \sigma_2}
  \left \vert\exp\left\{-\frac{1}{2 \sigma_2^2} (X_{i2} - t_{j2})^2 \right\}
  -
  \sum_{l = 0}^{L_n - 1} \frac{(-1)^{l} (X_{i2} - t_{j2})^{2l}}{(2 \sigma_2^2)^l l!}
  \right \vert \nonumber \\
  \leq\,&
  \frac{\gamma_n + 2}{2 \pi \sigma_1 \sigma_2} \sigma_1^q M_n^q \gamma_n. \label{eq:taylor1}
\end{align}

Next, by binomial expansion,
\begin{align*}
  &\frac{1}{n} \sum_{j = 1}^n \frac{(t_{j1} - X_{i1})^q}{2 \pi \sigma_1 \sigma_2} \prod_{d = 1}^2 \sum_{l = 0}^{L_n - 1} \frac{(-1)^l (X_{id} - t_{jd})^{2l}}{(2 \sigma_d^2)^l l!} = \frac{1}{2 \pi \sigma_1 \sigma_2 } \, \times\\
  & \sum_{\ell, \ell' = 0}^{L_n - 1} \frac{(-1)^{\ell} (-1)^{\ell'}}{(2 \sigma_1^2)^\ell \ell! (2 \sigma_2^2)^{\ell'} \ell'!}
  \sum_{m = 0}^{2 \ell + q} \sum_{m' = 0}^{2 \ell'} \binom{2 \ell + q}{m} (-X_{i1})^{2 \ell + q - m} \binom{2 \ell'}{m'} (-X_{i2})^{2 \ell' - m'}
  \sum_{j = 1}^n t_{j1}^m t_{j2}^{m'} n^{-1}.
\end{align*}
Lemma A.1 of \citet{ghosal2001entropies} can be used to show that for $K_n$ as in \eqref{eq:MLK}, there exist $\bmu \in \mathcal{U}(K_n)$ and $\bomega \in \mathcal{W}(K_n)$ \eqref{eq:sets} such that $\sum_{j = 1}^n t_{j1}^m t_{j2}^{m'}n = \sum_{k = 1}^{K_n} \mu_{k1}^m \mu_{k2}^{m'} \omega_k$ for every $0 \leq m \leq 2 \ell + q$ and $0 \leq m' \leq 2 \ell'$. Therefore
\begin{equation}
  \label{eq:lemmaA1}
  \begin{aligned}
    &\frac{1}{n} \sum_{j = 1}^n \frac{(t_{j1} - X_{i1})^q}{2 \pi \sigma_1 \sigma_2} \prod_{d = 1}^2 \sum_{l = 0}^{L_n - 1} \frac{(-1)^l (X_{id} - t_{jd})^{2l}}{(2 \sigma_d^2)^l l!}\\
    =\,&
    \sum_{k = 1}^{K_n} \frac{(\mu_{k1} - X_{i1})^q}{2 \pi \sigma_1 \sigma_2} \prod_{d = 1}^2 \sum_{l = 0}^{L_n - 1} \frac{(-1)^l (X_{id} - u_{kd})^{2l}}{(2 \sigma_d^2)^l l!} \omega_k.
  \end{aligned}
\end{equation}

Finally, by the same reasoning that led to \eqref{eq:taylor1},
\begin{align}
  &\left \vert
  \sum_{k = 1}^{K_n} \frac{(\mu_{k1} - X_{i1})^q}{2 \pi \sigma_1 \sigma_2}  \prod_{d = 1}^2 \sum_{l = 0}^{L_n - 1} \frac{(-1)^l (X_{id} - u_{kd})^{2l}}{(2 \sigma_d^2)^l l!} \omega_k
  -
  \sum_{k = 1}^{K_n} (u_{k1} - X_{i1})^q p(X_{i1}, X_{i2}; u_{k1}, u_{k2}) \omega_k
  \right \vert \nonumber\\
  \leq\,&
  \frac{\gamma_n + 2}{2 \pi \sigma_1 \sigma_2} \sigma_1^q M_n^q \gamma_n \label{eq:taylor2}
\end{align}
Combining \eqref{eq:taylor1}, \eqref{eq:lemmaA1}, and \eqref{eq:taylor2} concludes the proof.
$\square$

Next, for any $\gamma_n > 0$, define the discrete sets 
\begin{equation}
  \label{eq:discrete_sets}
  \begin{aligned}
    &\mathbb{U}(K_n) \subset \mathcal{U}(K_n) \mbox{ such that for any $\bmu \in \mathcal{U}(K_n)$,} \inf_{\bu \in \mathbb{U}(K_n)} \max_{k,d} \vert u_{kd} -  \mu_{kd} \vert \leq \gamma_n,\\
    &\mathbb{W}(K_n) \subset \mathcal{W}(K_n) \mbox{ such that for any $\bomega \in \mathcal{W}(K_n)$,} \inf_{\bw \in \mathbb{W}(K_n)} \sum_{k = 1}^{K_n} \vert w_k - \omega_k \vert \leq \gamma_n.
  \end{aligned}
\end{equation}
The set $\mathbb{U}(K_n)$ can be constructed by specifying a grid of equally-spaced points on the interval $[-C n^{1/4 - \eta}, C n^{1/4 - \eta}]$, where any two neighboring points are separated by at most a distance $\gamma_n$, for each of the $2 K_n$ dimensions of the vector $\bmu \in \mathcal{U}(K_n)$. Therefore the cardinality of the smallest possible $\mathbb{U}(K_n)$ satisfies
\begin{equation}
  \label{eq:covering_U}
  \vert \mathbb{U}(K_n) \vert \leq  \left( \lceil 2 C n^{1/4 - \eta} / \gamma_n \rceil + 1 \right)^{2 K_n}.
\end{equation}
Arguments from \citet{jiang2009general} and \citet{zhang2009generalized} imply that the cardinality of the smallest possible $\mathbb{W}(K_n)$ satifies
\begin{equation}
  \label{eq:covering_W}
  \vert \mathbb{W}(K_n) \vert \leq (2 / \gamma_n + 1)^{K_n}.
\end{equation}

\begin{lemma}
  \label{lem:K_to_grid}
  For any $(\bmu, \bomega) \in \mathcal{U}(K_n) \times \mathcal{W}(K_n)$ \eqref{eq:sets} and any integer $q > 0$, there exists $(\bu, \bw) \in \mathbb{U}(K_n) \times \mathbb{W}(K_n)$ \eqref{eq:discrete_sets} and a constant $D_q$ such that for each $(X_{i1}, X_{i2})$,
  \begin{align*}
    &\left \vert \sum_{k = 1}^{K_n} (\mu_{k1} - X_{i1})^q p(X_{i1}, X_{i2}; \mu_{k1}, \mu_{k2}) \omega_k -
    \sum_{k = 1}^{K_n} (u_{k1} - X_{i1})^q p(X_{i1}, X_{i2}; u_{k1}, u_{k2}) w_k \right \vert \\
    \leq\,&
    \left(M_n^q + \frac{D_q}{\sigma_1} + \frac{D_q M_n^q}{\sigma_2} \right)
    \frac{\sigma_1^q}{2 \pi \sigma_1 \sigma_2} \gamma_n.
  \end{align*}
\end{lemma}
\textit{Proof.} For any $(\bmu, \bomega) \in \mathcal{U}(K_n) \times \mathcal{W}(K_n)$, by construction of $\mathbb{W}(K_n)$ \eqref{eq:discrete_sets} there exists a $\bw \in \mathbb{W}(K_n)$ such that
\begin{equation}
  \label{eq:w}
  \left \vert \sum_{k = 1}^{K_n} (\mu_{k1} - X_{i1})^q p(X_{i1}, X_{i2}; \mu_{k1}, \mu_{k2}) (\omega_k - w_k) \right \vert
  \leq
  \frac{\sigma_1^q M_n^q}{2 \pi \sigma_1 \sigma_2} \gamma_n,
\end{equation}
and for any $\bu \in \mathcal{U}(K_n)$,
\begin{align*}
  &\left \vert \sum_{k = 1}^{K_n} \{ (\mu_{k1} - X_{i1})^q p(X_{i1}, X_{i2}; \mu_{k1}, \mu_{k2}) - (u_{k1} - X_{i1})^q p(X_{i1}, X_{i2}; u_{k1}, u_{k2}) \} w_k \right \vert\\
  \leq\,&
  \sum_{k = 1}^{K_n} \left \vert
  \frac{(\mu_{k1} - X_{i1})^q}{\sigma_1^q} \exp\left\{ -\frac{(X_{i1} - \mu_{k1})^2}{2 \sigma_1^2} \right\} - \frac{(u_{k1} - X_{i1})^q}{\sigma^q} \exp\left\{ -\frac{(X_{i1} - u_{k1})^2}{2 \sigma_1^2} \right\}
  \right \vert \frac{\sigma_1^q w_k}{2 \pi \sigma_1 \sigma_2} \,+\\
  & \sum_{k = 1}^{K_n} \left \vert
  \exp\left\{ -\frac{(X_{i2} - \mu_{k2})^2}{2 \sigma_2^2} \right\} - \exp\left\{ -\frac{(X_{i2} - u_{k2})^2}{ 2\sigma_2^2} \right\}
  \right \vert \frac{\sigma_1^q M_n^q w_k}{2 \pi \sigma_1 \sigma_2}.
\end{align*}
Define the function
\[
h_q(x) = x^q \exp(-x^2 / 2)
\]
for $q \geq 0$. Now if $\bu$ is also in $\mathbb{U}(K_n)$ \eqref{eq:discrete_sets}, then by construction
\begin{align*}
  &\left \vert \sum_{k = 1}^{K_n} \{ (\mu_{k1} - X_{i1})^q p(X_{i1}, X_{i2}; \mu_{k1}, \mu_{k2}) - (u_{k1} - X_{i1})^q p(X_{i1}, X_{i2}; u_{k1}, u_{k2}) \} w_k \right \vert\\
  \leq\,&
  \sum_{k = 1}^{K_n} \sup_x \vert h_q'(x) \vert
  \left \vert \frac{\mu_{k1} - X_{i1}}{\sigma_1} - \frac{u_{k1} - X_{i1}}{\sigma_1} \right \vert
  \frac{\sigma_1^q w_k}{2 \pi \sigma_1 \sigma_2} \,+\\
  & \sum_{k = 1}^{K_n} \sup_x \vert h_0'(x) \vert
  \left \vert \frac{\mu_{k2} - X_{i2}}{\sigma_2} - \frac{u_{k2} - X_{i2}}{\sigma_2} \right \vert
  \frac{\sigma_1^q M_n^q w_k}{2 \pi \sigma_1 \sigma_2} \\
  \leq\,&
  \sup_x \vert h_q'(x) \vert
  \frac{\sigma_1^q}{2 \pi \sigma_1^2 \sigma_2} \gamma_n +
  \sup_x \vert h_0'(x) \vert
  \frac{\sigma_1^q M_n^q}{2 \pi \sigma_1 \sigma_2^2} \gamma_n.
\end{align*}
Since
\begin{align*}
  h_q'(x) =\,& (q - x^2) x^{q - 1} \exp(-x^2 / 2), \\
  h_q''(x) =\,& \{ q (q - 1) - (2 q + 1) x^2 + x^4 \} x^{q - 2} \exp(-x^2 / 2),
\end{align*}
the first derivative of $\{ h_q'(x) \}^2$ is given by
\begin{align*}
  &
  \frac{d}{dx} \{ h_q'(x) \}^2
  =
  2 h_q'(x) h_q''(x)\\
  =\,&
  (q - x^2) \left\{ x^2 - \frac{2 q + 1 + (8 q + 1)^{1/2}}{2} \right\} \left\{ x^2 - \frac{2 q + 1 - (8 q + 1)^{1/2}}{2} \right\} x^{2 q - 3} \exp(-x^2 / 2).
\end{align*}
Therefore $\{ h_q'(x) \}^2$ has at most seven stationary points, is increasing for $x$ smaller than the smallest negative stationary point, and is decreasing for $x$ larger than the largest positive stationary point. Therefore the maximum value of $\vert h_q'(x) \vert$ can only occur at one of the seven stationary points, at which $\vert h_q'(x) \vert$ is finite. Therefore there is some constant $D_q$ such that
\begin{equation}
  \label{eq:u}
  \begin{aligned}
    &\left \vert \sum_{k = 1}^{K_n} \{ (\mu_{k1} - X_{i1})^q p(X_{i1}, X_{i2}; \mu_{k1}, \mu_{k2}) - (u_{k1} - X_{i1})^q p(X_{i1}, X_{i2}; u_{k1}, u_{k2}) \} w_k \right \vert\\
    \leq\,&
    D_q \left(
    \frac{\sigma_1^q}{2 \pi \sigma_1^2 \sigma_2} +
    \frac{\sigma_1^q M_n^q}{2 \pi \sigma_1 \sigma_2^2}
    \right) \gamma_n.
  \end{aligned}
\end{equation}
Combining \eqref{eq:w} and \eqref{eq:u} concludes the proof.
$\square$

For $K_n$ \eqref{eq:MLK} large enough, the result \eqref{eq:approxf}, that $f$ \eqref{eq:f} can be in a sense approximated by $g$ \eqref{eq:g}, can now be established using Lemmas \ref{lem:n_to_K} and \ref{lem:K_to_grid}. For each $(X_{i1}, X_{i2})$,
\begin{align}
  &\vert f(X_{i1}, X_{i2}, \theta_{i1}; \bt) - g(X_{i1}, X_{i2}, \theta_{i1}; \bu, \bw) \vert \leq \nonumber \\
  & \left \vert \frac{
    \sum_j^n (t_{j1} - X_{i1})^2 p(X_{i1}, X_{i2}; t_{j1}, t_{j2}) n^{-1}
  }{
    \rho n^{-1} + \sum_j^n p(X_{i1}, X_{i2}; t_1, t_2) n^{-1}
  }
  -
  \frac{
      \sum_k^{K_n} (u_{k1} - X_{i1})^2 p(X_{i1}, X_{i2}; u_{k1}, u_{k2}) w_k
    }{
      \rho n^{-1} + \sum_k^{K_n} p(X_{i1}, X_{i2}; u_{k1}, u_{k2}) w_k
  } \right \vert \,+ \label{eq:term1} \\
  &
  \sigma_1^2 \left \vert \frac{
    \sum_j^n p(X_{i1}, X_{i2}; t_{j1}, t_{j2}) n^{-1}
  }{
    \rho n^{-1} + \sum_j^n p(X_{i1}, X_{i2}; t_{j1}, t_{j2}) n^{-1}
  }
  -
  \frac{
    \sum_k^{K_n} p(X_{i1}, X_{i2}; u_{k1}, u_{k2})
  }{
    \rho n^{-1} + \sum_k^{K_n} p(X_{i1}, X_{i2}; u_{k1}, u_{k2}) w_k
  } \right \vert \,+ \label{eq:term2} \\
  &
  \left \vert \left\{
  \frac{
    \sum_j^n (t_{1j} - X_{i1}) p(X_{i1}, X_{i2}; t_{j1}, t_{j2}) n^{-1}
  }{
    \rho n^{-1} + \sum_j^n p(X_{i1}, X_{i2}; t_{j1}, t_{j2}) n^{-1}
  }
  \right\}^2
  -
  \left\{
  \frac{
    \sum_k^{K_n} (u_{k1} - X_{i1}) p(X_{i1}, X_{i2}; u_{k1}, u_{k2}) w_k
  }{
    \rho n^{-1} + \sum_k^{K_n} p(X_{i1}, X_{i2}; u_{k1}, u_{k2}) w_k
  }
  \right\}^2 \right \vert \,+ \label{eq:term3} \\
  &
  \vert X_{i1} - \theta_{i1} \vert
  \left \vert \frac{
    \sum_j^n (t_{1j} - X_{i1}) p(X_{i1}, X_{i2}; t_{j1}, t_{j2}) n^{-1}
  }{
    \rho n^{-1} + \sum_j^n p(X_{i1}, X_{i2}; t_{j1}, t_{j2}) n^{-1}
  }
  -
  \frac{
    \sum_k^{K_n} (u_{k1} - X_{i1}) p(X_{i1}, X_{i2}; u_{k1}, u_{k2}) w_k
  }{
    \rho n^{-1} + \sum_k^{K_n} p(X_{i1}, X_{i2}; u_{k1}, u_{k2}) w_k
  } \right \vert \label{eq:term4}.
\end{align}

The terms \eqref{eq:term1}--\eqref{eq:term4} can be bounded using \eqref{eq:jensen_rho} and Lemmas \ref{lem:n_to_K} and \ref{lem:K_to_grid}. For $L_n$ as defined in \eqref{eq:MLK} and
\begin{equation}
  \label{eq:K_n}
  K_n = 4 L_n^2 + 1,
\end{equation}
there exist $(\bu, \bw) \in \mathbb{U}(K_n) \times \mathbb{W}(K_n)$ \eqref{eq:discrete_sets} such that for any $\gamma_n \leq e^{-e}$,
\begin{align*}
  \eqref{eq:term1}
  \leq\,&
  \left \vert \frac{
    \sum_j^n (t_{j1} - X_{i1})^2 p(X_{i1}, X_{i2}; t_{j1}, t_{j2}) n^{-1}
  }{
    \rho n^{-1} + \sum_j^n p(X_{i1}, X_{i2}; t_1, t_2) n^{-1}
  }
  -
  \frac{
      \sum_k^{K_n} (u_{k1} - X_{i1})^2 p(X_{i1}, X_{i2}; u_{k1}, u_{k2}) w_k
    }{
      \rho n^{-1} + \sum_j^n p(X_{i1}, X_{i2}; t_{j1}, t_{j2}) n^{-1}
  } \right \vert \,+ \\
  &
  \left \vert \frac{
    \sum_k^{K_n} (u_{k1} - X_{i1})^2 p(X_{i1}, X_{i2}; u_{k1}, u_{k2}) w_k
  }{
    \rho n^{-1} + \sum_j^n p(X_{i1}, X_{i2}; t_{j1}, t_{j2}) n^{-1}
  }
  -
  \frac{
    \sum_k^{K_n} (u_{k1} - X_{i1})^2 p(X_{i1}, X_{i2}; u_{k1}, u_{k2}) w_k
  }{
    \rho n^{-1} + \sum_k^{K_n} p(X_{i1}, X_{i2}; u_{k1}, u_{k2}) w_k
  } \right \vert \\
  \leq\,&
  \frac{n}{\rho} \left \vert \sum_j^n (t_{j1} - X_{i1})^2 p(X_{i1}, X_{i2}; t_{j1}, t_{j2}) n^{-1} - \sum_k^{K_n} (u_{k1} - X_{i1})^2 p(X_{i1}, X_{i2}; u_{k1}, u_{k2}) w_k \right \vert \,+ \nonumber \\
  &
  \frac{n}{\rho}
  \frac{
    \sum_k^{K_n} (u_{k1} - X_{i1})^2 p(X_{i1}, X_{i2}; u_{k1}, u_{k2}) w_k
  }{
    \rho n^{-1} + \sum_k^{K_n} p(X_{i1}, X_{i2}; u_{k1}, u_{k2}) w_k
  } \, \times \\
  &
  \left \vert \sum_j^n p(X_{i1}, X_{i2}; t_{j1}, t_{j2}) n^{-1} - \sum_k^{K_n} p(X_{i1}, X_{i2}; u_{k1}, u_{k2}) w_k \right \vert,
\end{align*}
so by Lemmas \ref{lem:n_to_K} and \ref{lem:K_to_grid},
\begin{align*}
  \eqref{eq:term1}
  \leq\,&
  \frac{n}{\rho}
  \left\{
  \frac{(\gamma_n + 2) \sigma_1^2 M_n^2}{\pi \sigma_1 \sigma_2} +
  \left(M_n^2 + \frac{D_2}{\sigma_1} + \frac{D_2 M_n^2}{\sigma_2} \right) \frac{\sigma_1^2}{2 \pi \sigma_1 \sigma_2}
  \right\} \gamma_n \,+ \\
  &
  \frac{2 \sigma_1^2 n}{\rho^2} \log \frac{n}{2 \pi \sigma_1 \sigma_2}
  \left\{
  \frac{\gamma_n + 2}{\pi \sigma_1 \sigma_2} +
  \left(1 + \frac{D_0}{\sigma_1} + \frac{D_0}{\sigma_2} \right) \frac{1}{2 \pi \sigma_1 \sigma_2}
  \right\} \gamma_n.
\end{align*}
Similarly,
\begin{align*}
  \eqref{eq:term2}
  \leq\,&
  \sigma_1^2
  \left \vert \frac{
    \sum_j^n p(X_{i1}, X_{i2}; t_{j1}, t_{j2}) n^{-1}
  }{
    \rho n^{-1} + \sum_j^n p(X_{i1}, X_{i2}; t_{j1}, t_{j2}) n^{-1}
  }
  -
  \frac{
    \sum_k^{K_n} p(X_{i1}, X_{i2}; u_{k1}, u_{k2}) w_k
  }{
    \rho n^{-1} + \sum_j^n p(X_{i1}, X_{i2}; t_{j1}, t_{j2}) n^{-1}
  } \right \vert \,+ \\
  &
  \sigma_1^2
  \left \vert \frac{
    \sum_k^{K_n} p(X_{i1}, X_{i2}; u_{k1}, u_{k2}) w_k
  }{
    \rho n^{-1} + \sum_j^n p(X_{i1}, X_{i2}; t_{j1}, t_{j2}) n^{-1}
  }
  -
  \frac{
    \sum_k^{K_n} p(X_{i1}, X_{i2}; u_{k1}, u_{k2}) w_k
  }{
    \rho n^{-1} + \sum_k^{K_n} p(X_{i1}, X_{i2}; u_{k1}, u_{k2}) w_k
  } \right \vert \\
  \leq\,&
  \frac{\sigma_1^2 n}{\rho}
  \left \vert \sum_j^n p(X_{i1}, X_{i2}; t_{j1}, t_{j2}) n^{-1} - \sum_k^{K_n} p(X_{i1}, X_{i2}; u_{k1}, u_{k2}) w_k \right \vert \,+\\
  & \frac{\sigma_1^2 n}{\rho}
  \frac{
    \sum_k^{K_n} p(X_{i1}, X_{i2}; u_{k1}, u_{k2}) w_k
  }{
    \rho n^{-1} + \sum_k^{K_n} p(X_{i1}, X_{i2}; u_{k1}, u_{k2}) w_k
  } \, \times \\
  &
  \left \vert \sum_j^n p(X_{i1}, X_{i2}; t_{j1}, t_{j2}) n^{-1} - \sum_k^{K_n} p(X_{i1}, X_{i2}; u_{k1}, u_{k2}) w_k \right \vert \\
  \leq\,&
  \frac{2 \sigma_1^2 n}{\rho} 
  \left\{
  \frac{\gamma_n + 2}{\pi \sigma_1 \sigma_2} +
  \left(1 + \frac{D_0}{\sigma_1} + \frac{D_0}{\sigma_2} \right) \frac{1}{2 \pi \sigma_1 \sigma_2}
  \right\} \gamma_n.
\end{align*}
Next, using Jensen's inequality and the bound \eqref{eq:jensen_rho},
\begin{align*}
  \eqref{eq:term3}
  =\,&
  \left \vert
  \frac{
    \sum_j^n (t_{1j} - X_{i1}) p(X_{i1}, X_{i2}; t_{j1}, t_{j2}) n^{-1}
  }{
    \rho n^{-1} + \sum_j^n p(X_{i1}, X_{i2}; t_{j1}, t_{j2}) n^{-1}
  }
  +
  \frac{
    \sum_k^{K_n} (u_{k1} - X_{i1}) p(X_{i1}, X_{i2}; u_{k1}, u_{k2}) w_k
  }{
    \rho n^{-1} + \sum_k^{K_n} p(X_{i1}, X_{i2}; u_{k1}, u_{k2}) w_k
  }
  \right \vert \, \times \\
  &\left \vert
  \frac{
    \sum_j^n (t_{1j} - X_{i1}) p(X_{i1}, X_{i2}; t_{j1}, t_{j2}) n^{-1}
  }{
    \rho n^{-1} + \sum_j^n p(X_{i1}, X_{i2}; t_{j1}, t_{j2}) n^{-1}
  }
  -
  \frac{
    \sum_k^{K_n} (u_{k1} - X_{i1}) p(X_{i1}, X_{i2}; u_{k1}, u_{k2}) w_k
  }{
    \rho n^{-1} + \sum_k^{K_n} p(X_{i1}, X_{i2}; u_{k1}, u_{k2}) w_k
  }
  \right \vert \\
  \leq\,&
  2 \left( \frac{2 \sigma_1^2}{\rho} \log \frac{n}{2 \pi \sigma_1 \sigma_2} \right )^{1/2} \, \times\\
  \,&
  \left \vert
  \frac{
    \sum_j^n (t_{1j} - X_{i1}) p(X_{i1}, X_{i2}; t_{j1}, t_{j2}) n^{-1}
  }{
    \rho n^{-1} + \sum_j^n p(X_{i1}, X_{i2}; t_{j1}, t_{j2}) n^{-1}
  }
  -
  \frac{
    \sum_k^{K_n} (u_{k1} - X_{i1}) p(X_{i1}, X_{i2}; u_{k1}, u_{k2}) w_k
  }{
    \rho n^{-1} + \sum_k^{K_n} p(X_{i1}, X_{i2}; u_{k1}, u_{k2}) w_k
  }
  \right \vert.
\end{align*}
By similar reasoning as in the bounding of \eqref{eq:term1}, it follows that
\begin{align*}
  \eqref{eq:term3}
  \leq\,&
  2 \left( \frac{2 \sigma_1^2}{\rho} \log \frac{n}{2 \pi \sigma_1 \sigma_2} \right )^{1/2}
  \frac{n}{\rho}
  \left\{
  \frac{(\gamma_n + 2) \sigma_1 M_n}{\pi \sigma_1 \sigma_2} +
  \left(M_n + \frac{D_1}{\sigma_1} + \frac{D_1 M_n}{\sigma_2} \right) \frac{\sigma_1^1}{2 \pi \sigma_1 \sigma_2}
  \right\} \gamma_n \,+ \\
  &\frac{4 \sigma_1^2 n}{\rho^2} \log \frac{n}{2 \pi \sigma_1 \sigma_2}
  \left\{
  \frac{\gamma_n + 2}{\pi \sigma_1 \sigma_2} +
  \left(1 + \frac{D_0}{\sigma_1} + \frac{D_0}{\sigma_2} \right) \frac{1}{2 \pi \sigma_1 \sigma_2}
  \right\} \gamma_n,
\end{align*}
and that
\begin{align*}
  \eqref{eq:term4}
  \leq\,&
  \frac{\sigma_1 M_n n}{\rho}
  \left\{
  \frac{(\gamma_n + 2) \sigma_1 M_n}{\pi \sigma_1 \sigma_2} +
  \left(M_n + \frac{D_1}{\sigma_1} + \frac{D_1 M_n}{\sigma_2} \right) \frac{\sigma_1^1}{2 \pi \sigma_1 \sigma_2}
  \right\} \gamma_n \,+ \\
  &\frac{\sigma_1 M_n n}{\rho} \left( \frac{2 \sigma_1^2}{\rho} \log \frac{n}{2 \pi \sigma_1 \sigma_2} \right)^{1/2}
  \left\{
  \frac{\gamma_n + 2}{\pi \sigma_1 \sigma_2} +
  \left(1 + \frac{D_0}{\sigma_1} + \frac{D_0}{\sigma_2} \right) \frac{1}{2 \pi \sigma_1 \sigma_2}
  \right\} \gamma_n.
\end{align*}

Finally, choose
\begin{equation}
  \label{eq:gamma_n}
  \gamma_n = \{ n (M_n + 1)^2 \log^2 n \}^{-1},
\end{equation}
so that as long as $n \log^2 n > e^e$, $\gamma_n \leq n^{-e}$ and Lemmas \ref{lem:n_to_K} and \ref{lem:K_to_grid} can be applied to obtain the bounds on \eqref{eq:term1}--\eqref{eq:term4}. These bounds then imply that for $C_n$ as in \eqref{eq:C_n}, $M_n$ and $L_n$ as in \eqref{eq:MLK}, $K_n$ as in \eqref{eq:K_n}, and $\mathbb{U}(K_n)$ and $\mathbb{W}(K_n)$ constructed as in \eqref{eq:discrete_sets}, for any $\bt \in \mathcal{T}$ \eqref{eq:T} there exists some $(\bu, \bw) \in \mathbb{U}(K_n) \times \mathbb{W}(K_n)$ such that conditional on the $(X_{i1}, X_{i2})$,
\[
\frac{1}{n} \sum_{i = 1}^n \vert f(X_{i1}, X_{i2}, \theta_{i1}; \bt) - g(X_{i1}, X_{i2}, \theta_{i1}; \bu, \bw) \vert 
\leq
O\left\{ n (M_n + 1)^2 \log n \right\} \gamma_n.
\]
With $\gamma_n$ as in \eqref{eq:gamma_n}, for any $\epsilon > 0$ there exists a sufficiently large $n$ such that the approximation result \eqref{eq:approxf} holds. The cardinality bounds \eqref{eq:covering_U} and \eqref{eq:covering_W} imply that the $\epsilon$-covering number $\mathcal{N}(\mathcal{T})$ of the $f(x_1, x_2, \mu; \bt)$ indexed by $\bt \in \mathcal{T}$ obeys
\begin{align*}
  \log \mathcal{N}(\mathcal{T})
  \leq\,&
  2 K_n \log (2 C n^{1/4 - \eta} / \gamma_n + 1) + K_n \log (2 / \gamma_n + 1) \\
  \leq\,&
  K_n \log C_1 + K_n \log (n^{1/2 - 2 \eta} / \gamma_n^3 ),
\end{align*}
where $C$ and $\eta$ are defined in Assumption \ref{a:theta} and $C_1$ and $C_2$ are positive constants. Since $K_n = 4 L_n^2 + 1$ from \eqref{eq:K_n} and $L_n = (1 + M_n^2 e / 2) \log (1 / \gamma_n) > 1$ from \eqref{eq:MLK}, $K_n \leq 5 L_n^2$ and there is a constant $C_2$ such that
\begin{equation}
  \label{eq:covering}
  \begin{aligned}
    \log \mathcal{N}(\mathcal{T})
    \leq\,&
    C_2 (M_n + 1)^4 \log^2 \{ n (M_n + 1)^2 \log^2 n \} \times \\
    & \left[ \log C_1 + \log \{ n^{1/2 - 2 \eta} n^3 (M_n + 1)^6 \log^6 n \} \right].
  \end{aligned}
\end{equation}

\subsection{Maximal inequality}
This section concludes the proof of Theorem \ref{thm:uniform} by showing \eqref{eq:trunc}. First, the supremum over the infinite set $\mathcal{T}$ can be converted to a maximum over the discrete set $\mathbb{U}(K_n) \times \mathbb{W}(K_n)$ constructed in \eqref{eq:discrete_sets}. For any $\bt \in \mathcal{T}$, using the approximation result \eqref{eq:approxf} there exists a $(\bu, \bw)$ inside a suitably constructed set $\mathbb{U}(K_n) \times \mathbb{W}(K_n)$ \eqref{eq:discrete_sets} such that for any $\epsilon > 0$, there exists a sufficiently large $n$ such that
\begin{align*}
  &\left \vert \frac{1}{n} \sum_i \varepsilon_i f(X_{i1}, X_{i2}, \theta_{i1}; \bt) I(\vert X_{i1} - \theta_{i1}\vert \leq C_n ) \right \vert \\
  \leq\,&
  \left \vert \frac{1}{n} \sum_i \varepsilon_i g(X_{i1}, X_{i2}, \theta_{i1}; \bu, \bw) I(\vert X_{i1} - \theta_{i1}\vert \leq C_n ) \right \vert + \epsilon,
\end{align*}
where $\varepsilon_i$ are Rademacher variables and $C_n = (2 \sigma_1^2 \log \log n)^{1/2}$ from \eqref{eq:C_n}. Since $\mathbb{U}(K_n) \times \mathbb{W}(K_n)$ is discrete, it follows that
\begin{align*}
  & E \sup_{\bt \in \mathcal{T}} \left \vert \frac{1}{n} \sum_i \varepsilon_i f(X_{i1}, X_{i2}, \theta_{i1}; \bt) I(\vert X_{i1} - \theta_{i1}\vert \leq C_n ) \right \vert \\
  \leq\,&
  E \max_{(\bu, \bw) \in \mathbb{U}(K_n) \times \mathbb{W}(K_n)} \left \vert \frac{1}{n} \sum_i \varepsilon_i g(X_{i1}, X_{i2}, \theta_{i1}; \bu, \bw) I(\vert X_{i1} - \theta_{i1}\vert \leq C_n ) \right \vert + \epsilon.
\end{align*}

To bound the expectation of the maximum over $\mathbb{U}(K_n) \times \mathbb{W}(K_n)$, define the convex function $\psi(x) = \exp(x^2) - 1$. Using Lemma 2.2.2 of \citet{vandervaart1996weak}, it can be shown that there is a constant $A$ such that
\begin{align*}
  &E_\varepsilon \max_{(\bu, \bw) \in \mathbb{U}(K_n) \times \mathbb{W}(K_n)} \left \vert \frac{1}{n} \sum_i \varepsilon_i g(X_{i1}, X_{i2}, \theta_{i1}; \bu, \bw) I(\vert X_{i1} - \theta_{i1}\vert \leq C_n ) \right \vert \\
  \leq\,&
  A \{1 + \log \mathcal{N}(\mathcal{T})\}^{1/2} \max_{(\bu, \bw) \in \mathbb{U}(K_n) \times \mathbb{W}(K_n)} \left \Vert \frac{1}{n} \sum_i \varepsilon_i g(X_{i1}, X_{i2}, \theta_{i1}; \bu, \bw) I(\vert X_{i1} - \theta_{i1}\vert \leq C_n ) \right \Vert_\psi,
\end{align*}
where $E_\varepsilon$ is the expectation over the $\varepsilon_i$ conditional on the $(X_{i1}, X_{i2})$ and $\Vert X \Vert_\psi$ is the Orlicz norm of the random variable $X$ with respect to the function $\psi(x)$. It is straightforward to show that $\vert g(x_1, x_2, \mu; \bu, \bw) \vert \leq F_n(x_1, \mu)$ for the function $F_n$ defined in \eqref{eq:F_n}, so by Hoeffding's inequality and Lemma 2.2.7 of \citet{vandervaart1996weak},
\begin{align*}
  &\left \Vert \frac{1}{n} \sum_i \varepsilon_i g(X_{i1}, X_{i2}, \theta_{i1}; \bu, \bw) I(\vert X_{i1} - \theta_{i1}\vert \leq C_n ) \right \Vert_\psi\\
  \leq\,&
  \left( \frac{6}{n} \right)^{1/2}
  \left( \sigma_1^2 + \frac{4 \sigma_1^2}{\rho} \log \frac{n}{2 \pi \sigma_1 \sigma_2} + \frac{2^{1/2} \sigma_1}{\rho^{1/2}} C_n \log^{1/2} \frac{n}{2 \pi \sigma_1 \sigma_2} \right)
  =
  O\left( \frac{\log n}{n^{1/2}} \right)
\end{align*}
for any $(\bu, \bw) \in \mathbb{U}(K_n) \times \mathbb{W}(K_n)$. Therefore, using Jensen's inequality,
\[
E \sup_{\bt \in \mathcal{T}} \left \vert \frac{1}{n} \sum_i \varepsilon_i f(X_{i1}, X_{i2}, \theta_{i1}; \bt) I(\vert X_{i1} - \theta_{i1}\vert \leq C_n ) \right \vert
\leq
\left[ A \{1 + E \log \mathcal{N}(\mathcal{T}) \}^{1/2} \right] O\left( \frac{\log n}{n^{1/2}} \right).
\]

To show \eqref{eq:trunc}, it remains to show that $E \log \mathcal{N}(\mathcal{T}) = o(n / \log^2 n)$, where the expectation is now over the $(X_{i1}, X_{i2})$. From \eqref{eq:covering} and the inequality $\log(x) \leq m x^{1 / m}$ for any $x > 0$ and integer $m > 0$, for $n$ sufficiently large there is some constant $A_2$ such that
\begin{align*}
  E \log \mathcal{N}(\mathcal{T})
  \leq\,&
  A_2 ( n^{11 / 2m - 2 \eta / m} \log^{10 / m} n ) E \{ (M_n + 1)^{4 + 10 / m} \}.
\end{align*}
By the definition of $M_n$ \eqref{eq:MLK} and the bounds on $\theta_{id}$ from Assumption \ref{a:theta},
\[
M_n
=
\max_{j,d} \frac{\vert X_{jd} \vert + C n^{1/4 - \eta}}{\sigma_d}
\leq
\max_{j,d} \left \vert \frac{X_{jd} - \theta_{jd}}{\sigma_d} \right \vert + \frac{2 C n^{1/4 - \eta}}{\sigma_1 \wedge \sigma_2}.
\]
Therefore letting $Z_i$, $i = 1, \ldots, 2n$ denote independent standard normal random variables,
\begin{align*}
  E \{ (M_n + 1)^{4 + 10 / m} \}
  \leq\,&
  E \left \{ (M_n + 1)^{4 + 10 / m} I \left( \max_{j,d} \left \vert \frac{X_{jd} - \theta_{jd}}{\sigma_d} \right \vert \leq \frac{2 C n^{1/4 - \eta}}{\sigma_1 \wedge \sigma_2} + 1 \right) \right \} \, + \\
  &E \left \{ (M_n + 1)^{4 + 10 / m} I \left( \max_{j,d} \left \vert \frac{X_{jd} - \theta_{jd}}{\sigma_d} \right \vert > \frac{2 C n^{1/4 - \eta}}{\sigma_1 \wedge \sigma_2} + 1 \right) \right \} \\
  \leq\,&
  \left( \frac{4 C n^{1/4 - \eta}}{\sigma_1 \wedge \sigma_2} + 2 \right)^{4 + 10/m}
  +
  2^{4 + 10/m} E ( \max_{i = 1, \ldots, 2 n} \vert Z_i \vert^{4 + 10/m} ).
\end{align*}
Now define the convex function $\psi(x) = x^{p / (4 + 10 / m)}$ for $p \geq 4 + 10/m$. Then by Lemma 2.2.2 of \citet{vandervaart1996weak}, there is some constant $A_3$ such that
\[
E( \max_i \vert Z_i \vert^{4 + 10/m} )
\leq
\psi^{-1}(1) \left \Vert \max_{i = 1, \ldots, 2 n} \vert Z_i \vert^{4 + 10/m} \right \Vert_\psi
\leq
A_3 (2 n)^{(4 + 10/m) / p} \max_{i = 1, \ldots, 2 n} \left \Vert \vert Z_i \vert^{4 + 10/m} \vert \right \Vert_\psi,
\]
where $\Vert X \Vert_\psi$ is the Orlicz norm of $X$ with respect to the function $\psi(x)$. But
\[
\left \Vert \vert Z_i \vert^{4 + 10/m} \vert \right \Vert_\psi
=
E( \vert Z_i \vert^p )^{ (4 + 10 / m) / p },
\]
which is a constant that does not depend on $n$, so choosing $p$ such that $p \geq 5$ and $1 / p \leq 1/4 - \eta$ gives
\[
E\{ (M_n + 1)^{4 + 10 / m} \} \leq O(n^{1 - 4 \eta + 10 / (4 m) - 12 \eta / m} ).
\]
Then
\begin{align*}
  E \log \mathcal{N}(\mathcal{T})
  \leq\,&
  O( n^{1 - 4 \eta + 32 / (4 m) - 12 \eta / m} \log^{10 / m} n ).
\end{align*}
Choosing $m$ large enough such that $4 \eta - 32 / (4 m) + 10 \eta / m > 0$ implies that $E \log \mathcal{N}(\mathcal{T}) = o(n / log^2 n)$, concluding the proof.
$\square$

\section{Proof of Theorem \ref{thm:opt_risk}}
Let $\btheta = (\theta_{11}, \ldots, \theta_{n1}, \theta_{12}, \ldots, \theta_{n2})$. Then when $\bt = \btheta$, the estimator $\bdelta^{\bt}_\rho$ \eqref{eq:delta^t} equals the oracle regularized estimator $\bdelta^\star_\rho$. Therefore with $\textsc{sure}(\bt)$ defined as in \eqref{eq:sure},
\begin{align*}
  E \ell_n(\hat{\bt}) - R_n(\btheta, \bdelta^\star)
  =\,&
  E \ell_n(\hat{\bt}) - E \textsc{sure}(\hat{\bt}) + E \textsc{sure}(\hat{\bt}) - E \textsc{sure}(\theta) + E \textsc{sure}(\theta) - R_n(\btheta, \bdelta^\star) \\
  \leq\,&
  \vert E \{ \ell_n(\hat{\bt}) - \textsc{sure}(\hat{\bt}) \} \vert + R_n(\btheta, \bdelta^\star_\rho) - R_n(\btheta, \bdelta^\star) \\
  \leq\,&
  E \vert \ell_n(\hat{\bt}) - \textsc{sure}(\hat{\bt}) \vert + o(1),
\end{align*}
where the second line follows because $E \textsc{sure}(\hat{\bt}) \leq E \textsc{sure}(\theta)$, since $\hat{\bt}$ \eqref{eq:proposed} is defined to minimize $\textsc{sure}(\bt)$, and the third line follows by Theorem \ref{thm:regularized}. Finally, since $\hat{\bt} \in \mathcal{T}$ by construction, by Theorem \ref{thm:uniform},
\[
E \vert \ell_n(\hat{\bt}) - \textsc{sure}(\hat{\bt}) \vert
\leq
E \sup_{\bt \in \mathcal{T}} \vert \ell_n(\bt) - \textsc{sure}(\bt) \vert = o(1).
\]

\end{document}